\documentclass{article}

\usepackage{arxiv}

\usepackage[utf8]{inputenc} 
\usepackage[T1]{fontenc}    
\usepackage{url}            
\usepackage{booktabs}       
\usepackage{amsfonts}       
\usepackage{nicefrac}       
\usepackage{microtype}      
\usepackage{lipsum}		
\usepackage{graphicx}
\usepackage{doi}

\usepackage{amsmath}
\usepackage{amsthm}
\usepackage{amssymb}
\usepackage{amsthm} 
\usepackage{ascmac}
\usepackage{bm}
\usepackage{cases}
\usepackage{comment} 
\usepackage{color} 
\usepackage{enumerate}
\usepackage{float} 
\usepackage{graphicx} 
\usepackage{mathptmx} 
\usepackage{mathrsfs}
\usepackage{url}
\usepackage{color}
\usepackage{ulem}
\usepackage{algorithmic}

\usepackage[ruled]{algorithm2e} 

\SetAlFnt{\small}
\SetAlCapFnt{\small}
\SetAlCapNameFnt{\small}
\SetAlCapHSkip{0pt}
\IncMargin{-\parindent}

\theoremstyle{definition} 

\newtheorem{theorem}{Theorem}[section]

\newtheorem{lemma}[theorem]{Lemma}

\newtheorem{corollary}[theorem]{Corollary}

\newcommand{\remkcmd}{\\ {\bf \remkcmd} \ }

\newtheorem{assump}{Assumption}

\def\max{\mathop{\rm max}}
\def\min{\mathop{\rm min}}

\title{Estimating heterogeneous treatment effects by W-MCM based on Robust reduced rank regression}

\author{ \href{https://orcid.org/0000-0000-0000-0000}{\includegraphics[scale=0.06]{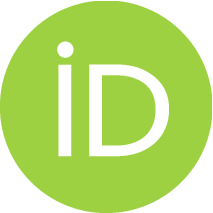}\hspace{1mm}Ryoma Hieda$^*$}\\
	Graduate School of Culture and Information Science\\
	Doshisha University, Japan\\
	\texttt{cishieda@gmail.com} \\
	\And
	\href{https://orcid.org/0000-0000-0000-0000}{\includegraphics[scale=0.06]{orcid.eps}\hspace{1mm}Shintaro Yuki$^*$}\\
	Graduate School of Culture and Information Science\\
	Doshisha University, Japan\\
	\texttt{} \\
    \And
	\href{https://orcid.org/0000-0000-0000-0000}{\includegraphics[scale=0.06]{orcid.eps}\hspace{1mm}Kensuke Tanioka} \\
	Department of Biomedical Sciences and
Informatics\\
	Doshisha University, Japan\\
	\texttt{} \\
 \And
	\href{https://orcid.org/0000-0000-0000-0000}{\includegraphics[scale=0.06]{orcid.eps}\hspace{1mm}Hiroshi Yadohisa}\\
	Department of Culture and Information Science\\
	Doshisha University, Japan\\
	\texttt{} \\
}

\date{}

\renewcommand{\shorttitle}{\textit{arXiv} Template}

\hypersetup{
pdftitle={A template for the arxiv style},
pdfsubject={q-bio.NC, q-bio.QM},
pdfauthor={David S.~Hippocampus, Elias D.~Striatum},
pdfkeywords={First keyword, Second keyword, More},
}

\begin{document}
\maketitle
\begin{gather*}
*\;\mathrm{ These\;two\;authors\;contributed\;equally\;to\;this\;work.}
\end{gather*}
\\
\begin{abstract}
Recently, from the personalized medicine perspective, there has been an increased demand to identify subgroups of subjects for whom treatment is effective. 
Consequently, the estimation of heterogeneous treatment effects (HTE) has been attracting attention.
While various estimation methods have been developed for a single outcome, there are still limited approaches for estimating HTE for multiple outcomes. 
Accurately estimating HTE remains a challenge especially for datasets where there is a high correlation between outcomes or the presence of outliers.
Therefore, this study proposes a method that uses a robust reduced-rank regression framework to estimate treatment effects and identify effective subgroups. 
This approach allows the consideration of correlations between treatment effects and the estimation of treatment effects with an accurate low-rank structure.
It also provides robust estimates for outliers. 
This study demonstrates that, when treatment effects are estimated using the reduced rank regression framework with an appropriate rank, the expected value of the estimator equals the treatment effect. 
Finally, we illustrate the effectiveness and interpretability of the proposed method through simulations and real data examples.
\end{abstract}
\section{Introduction}
\quad Recently, in medicine, there has been an increasing demand for optimizing the treatment effects according to individual characteristics, called personalized medicine.
Since treatment effects can vary across patients (Dahabreh et al. 2016), personalized medicine primarily entails the estimation and interpretation of heterogeneous treatment effects (HTE).
Here, we focus on explanatory analyses.
In this context, there are various methods for estimating HTE (Lunceford and Davidian 2004, Imbens and Rubin 2015, Athey and Imbens 2017, Twisk et al.
2018).
However, many of these have primarily focused on improving prediction accuracy, making it difficult to interpret subgroups through covariates (Wager and Athey 2018).
In a practical situation, physicians and researchers need interpretable methods.
Furthermore, these methods need to be designed to estimate treatment effects on a single outcome. 
However, the estimation of HTE is considered exploratory data analysis (Lipkovich et al. 2017), for which it is natural to handle multiple outcomes.
Therefore, we focus on estimating the effects of treatment on multiple outcomes.
There are few studies on the estimation of the treatment effects on multiple outcomes (Siriwardhana and Kulasekera 2020, Guo et al. 2021, Kulasekera and Siriwardhana
2022, Yuki et al. 2023).
Among them, we adopted the weighting method based on the framework of multiple outcomes (Yuki et al. 2023), which is in turn based on the single outcome method of Chen et al. (2017).
This is because it is easy to visually interpret covariates that contribute to the treatment effects through a path diagram.
Additionally, this method can be used in both randomized controlled trials (RCTs) and observational studies.
However, this method does not consider the following methodological and theoretical considerations.

\quad From a methodological perspective, the handling of outliers has not been considered.
In clinical trial data, outliers may be present in outcomes, which makes it crucial to determine how to handle them when estimating HTE.
Indeed, Li et al. (2023) refer to the need to deal with outliers and propose a method to robustly estimate HTE for a single outcome.
However, when handling multiple outcomes, there is a possibility that estimated treatment effects may be influenced by outliers.
Therefore, we propose a method using the M-estimation framework of Welsh (1989) and Liu et al. (2022).

\quad From a theoretical perspective, it has not been yet demonstrated that treatment effects can be estimated when the true regression coefficient matrix for treatment effects has a low-rank structure.
Yuki et al. (2023) prove that, under the implicit assumption that a regression coefficient matrix for estimating treatment effects is full-rank, the expected value of the estimated treatment effects becomes the conditional average treatment effects (CATE).
However, when dealing with multiple outcomes, a high correlation between outcomes may be correlated with treatment effects; as such, the above regression coefficient matrix may have a low-rank structure.
Therefore, we prove that the expected value of the treatment effect estimated by a weighting method based on a reduced-rank regression is the CATE.
Chen et al. (2017) and Yuki et al. (2023) prove the property of the estimator using the framework of the convex function.
However, in the reduced-rank regression framework, the parameter estimation problem can be bi-convex with respect to parameters.
Therefore, the existing methods do not cover this problem.
In this study, we theoretically prove that the expected value of the weighting method equals the true treatment effects.

\quad The remainder of paper is organized as follows.
In Section 2, we describe the notation, assumptions, model equations, objective function, and algorithm of the proposed method.
In Section 3, we present the properties of the estimator that minimize the expected value of the objective function for the weighting method based on reduced-rank regression.
In Section 4, we perform numerical simulations of the proposed method and present comparison methods.
In Section 5, we explain the application of the proposed method to real data.
In Section 6, we conclude the paper.
The proof of the theorem for the weighting method based on reduced-rank regression in Section 3 and detailed simulation results can be found in the Supplementary Material.

\section{Proposed Method}
\quad Here, we first introduce the basic notations and then describe the formula and assumptions of the proposed method. Subsequently, we present the objective function and the updated formula and algorithm.
We adopt the weighting method from among the various methods for estimating treatment effects (Guo et al. 2021, Yuki et al. 2023).
In this method, the accuracy of estimated treatment effects becomes worse when 
latent multiple treatment effects have a low-rank structure or outliers are present in outcomes.
Therefore, rank constraints are imposed on the parameters to be estimated and parameters for adjusting outliers are introduced. 
The estimation algorithm of the proposed method utilizes the alternating least squares method (Young et al. 1976).
\subsection{Models and assumptions}
\quad We first present the notations for the proposed model. 
Let $\bm{X}_{\rm{(r.v.)}}=(\bm{x}_{1\rm{(r.v.)}},\bm{x}_{2\rm{(r.v.)}},\ldots,\\
\bm{x}_{n\rm{(r.v.)}})^{\top}\in\mathbb{R}^{n\times p}$ be the matrix of random variables corresponding to the covariates with $n$ subjects and $p$ variables, where $\bm{x}_{i\rm{(r.v.)}}\in\mathbb{R}^{p}$ for subject $i$, 
$\bm{X}_{\rm{(o.b.)}}=(\bm{x}_{1\rm{(o.b.)}},\bm{x}_{2\rm{(o.b.)}},\ldots,\bm{x}_{n\rm{(o.b.)}})^{\top}\in\mathbb{R}^{n\times p}$ is the observed value of $\bm{X}_{\rm{(r.v.)}}$, 
 $\bm{Y}=(Y_{1}, Y_{2},\ldots, Y_{n})^{\top}\in\mathbb{R}^{n\times q}$ be the matrix of the random variable corresponding to the multiple continuous outcomes with $n$ subjects and $q$ outcomes, and
$\bm{Y}_{\rm{(o.b.)}}=(Y_{1\rm{(o.b.)}}, Y_{2\rm{(o.b.)}},\ldots, Y_{n\rm{(o.b.)}})^{\top}\in\mathbb{R}^{n\times p}$ be the observed value of $\bm{Y}$.
Let $\bm{T}\in\mathbb{R}^{n\times n}={\rm{diag}}(T_{1},T_{2},\ldots,T_{n})$ be the diagonal matrix of random variable, whose diagonal element $T_{i}$ is defined as follows:
\begin{equation*}
\centering
T_{i}= \left \{
\begin{array}{l}
1\quad \mbox{(If subject}\ i \ \mbox{is assigned to the treatment group)} \\
-1\quad \mbox{(If subject}\ i \ \mbox{is assigned to the control group)},
\end{array}
\right.
\end{equation*}
$\bm{T}_{\rm(o.b.)}\in\mathbb{R}^{n\times n}={\rm{diag}}(T_{1\rm(o.b.)},T_{2\rm(o.b.)},\ldots,T_{n\rm(o.b.)})$
be the observed value of $\bm{T}$, and $\bm{E}=(\varepsilon_{1},\varepsilon_{2}\ldots,\varepsilon_{n})^{\top}\\\in\mathbb{R}^{n\times q}$ be the matrix of the random variable corresponding to the error; we also assume $\mathbb{E}[\bm{E}]=\bm{O}$ where all elements of matrix $\bm{O}\in\mathbb{R}^{n\times q}$ are zero. Furthermore, we assume that data $\{Y_{i},T_{i},\bm{x}_{i{\rm{(r.v.)}}}\}_{i=1}^{n}$ are independent
and identically distributed.

\quad We introduce a simple multivariate regression model for multiple outcomes. The model of the outcomes for subject $i$ can be expressed as follows:
\begin{align}
\label{y_model}
{Y}_{i}={b}(\bm{x}_{i\rm{(r.v.)}})+\frac{T_{i}}{2}\bm{\varGamma}_{\rm{(true)}}^{\top}\bm{x}_{i\rm{(r.v.)}}+\bm{\varepsilon}_{i} ,
\end{align}
where $\bm{\varGamma}_{\rm{(true)}}\in\mathbb{R}^{p \times q}$ is the true regression coefficient matrix for treatment effects and $b(\bm{x}_{i\rm{(r.v.)}}): \mathbb{R}^{p}\rightarrow\mathbb{R}^{q}$ is a function for the true main effects term.
This model is equivalent to those of Guo et al.(2021) and Yuki et al. (2023). 
In addition, we define the notation and model formula of potential outcomes (Rubin 1974, Splawa-Neyman et al. 1990). Based on that, we formulate multivariate potential outcomes.

Let $Y_{i}^{(1)}\in\mathbb{R}^{q}$ be ${Y}_{i}$ if subject $i$ is assigned to the treatment group and ${Y}_{i}^{(-1)}\in\mathbb{R}^{q}$ be ${Y}_{i}$ if subject $i$ is assigned to the control group. 
The model formulae for these potential outcomes of subject $i$, $Y_i^{(1)}$ and $Y_i^{(-1)}$, are defined as Eqs. (\ref{senzaiout_model1}) and (\ref{senzaiout_model-1}), respectively. 
\begin{equation}
\label{senzaiout_model1}
{Y}_{i}^{(1)}={b}(\bm{x}_{i{\rm{(r.v.)}}})+\frac{{1}}{2}\bm{\varGamma}_{\rm{(true)}}^{\top}\bm{x}_{i{\rm{(r.v.)}}}+{\varepsilon}_{i}, \\
\end{equation}
\begin{equation}
\label{senzaiout_model-1}
{Y}_{i}^{(-1)}={b}(\bm{x}_{i{\rm{(r.v.)}}})-\frac{{1}}{2}\bm{\varGamma}_{\rm{(true)}}^{\top}\bm{x}_{i{\rm{(r.v.)}}}+{\varepsilon}_{i}.
\end{equation}
In this study, we focused on the CATE for multivariate continuous outcomes denoted as $\tau$.
Given Eqs. \ref{senzaiout_model1} and \ref{senzaiout_model-1}, $\tau$ is expressed as a linear combination of explanatory variables:
\begin{align*}
\tau{(\bm{x}_{i\rm{(o.b.)}})}&=E_{{Y}_{i}^{(1)}, {Y}_{i}^{(-1)}}[{Y}_{i}^{(1)}-{Y}_{i}^{(-1)}|\bm{x}_{i\rm{(r.v.)}}=\bm{x}_{i\rm{(o.b.)}}]\\
&=E_{{Y}}[{Y}_{i}|\bm{x}_{i\rm{(r.v.)}}=\bm{x}_{i\rm{(o.b.)}},{T}_{i}=1]-E_{{Y}}\left[{Y}_{i}|\bm{x}_{\rm{(r.v.)}}=\bm{x}_{i\rm{(o.b.)}}, {T}_{i}=-1\right]\\
&=\bm{\varGamma}_{\rm{(true)}}^{\top}\bm{x}_{i\rm{(o.b)}}.
\end{align*}
Furthermore, the main effects are expressed as:
${b}(\bm{x}_{i{\rm{(o.b.)}}})=\frac{1}{2}(E[{Y}_{i}^{(1)}|\bm{x}_{i\rm{(r.v)}}=\bm{x}_{i\rm{(o.b)}}]+E[{Y}^{(-1)}|\\\bm{x}_{i\rm{(r.v)}}=\bm{x}_{i\rm{(o.b)}}])$.
We assume that true treatment effects have correlations and, in this situation, the assumption of $\bm{\varGamma}_{\rm{(true)}}$ with low rank is natural. Therefore, we deal with the model as such in Section \ref{weighting method for}.

\quad We also assume the following four conditions to estimate CATE: 
\begin{assump}[SUTVA (Stable Unit Treatment Value Assumption)]
\rm{The treatment assigned to one subject does not affect the outcomes of the other subjects and the treatment for subjects} $i$ and $j$ assigned to the same group is the same (Imbens and Rubin 2015).
\end{assump}
\begin{assump}[Ignorability]
\rm{Only the covariates affect the treatment assignment, that is,}
$\{Y_{i}^{(1)},Y_{i}^{(-1)} \perp T_{i}|\bm{x}_{i\rm{(r.v.)}}=\bm{x}_{i\rm{(o.b.)}}\}$ \rm{(Rosenbaum and Rubin 1983)}.
\end{assump}

\begin{assump}[Positivity]\rm{Observed values exist for both treatment and control groups under the condition of the observed covariates,} $\bm{x}_{i\rm{(o.b.)}}$, that is, 
 $0<{\rm{P}}(T_{i}=1|\bm{x}_{i})<1$ \rm{(Rosenbaum and Rubin 1983)}.

\end{assump}
\begin{assump}[Conditional Independence Error]
Error and treatment assignment are independent under conditioning on covariates, that is, 
$\varepsilon_{i}\perp T_{i}|\bm{x}_{i}$ \rm{(Li et al. 2023)}.
\end{assump}

\subsection{Weighting method for multiple outcomes in the robust estimation framework }
\label{weighting method for}
\quad Here, we describe the objective function of the proposed method, referred to as Weighting Modified Covariates Method based on Robust reduced rank regression (WMCMR4). It is a weighting method based on the sparse reduced rank regression framework (Liu et al. 2022), which enables rank-constrained and robust estimation of the regression coefficient matrix for the treatment effects even when the outcomes contain outliers.

\quad We assume that:
\begin{align*}
    {\rm{rank}}(\bm{\varGamma})=r\leq{\rm{min}}(p,q),
\end{align*}
where $\bm{\varGamma}\in\mathbb{R}^{p\times q}$ is the estimating regression coefficient matrix for treatment effects. 
The set of $r$ linear combinations of the treatment effects can be interpreted as latent factors and we expected to account for correlations among $q$ multiple treatment effects through $r$ common latent factors.
In this case, $\bm{\varGamma}$ can be expressed as a product of low dimensional matrices of full rank (Reinsel et al. 2022). Specifically, $\bm{\varGamma}$ can be written as:
\begin{align}
\label{gamma_decom}
    \bm{\varGamma}=\bm{W}\bm{V}^{\top},
\end{align}
where $\bm{W}\in\mathbb{R}^{n\times r}$ is the matrix and 
$\bm{V}\in\mathbb{R}^{q\times r}$ is the orthogonal matrix. 
We can interpret $\bm{W}$ as a mapping that transforms $\bm{T}\bm{X}_{\rm{(r.v.)}}/2$ into $r$ latent variables
and $\bm{V}$ as a mapping that transforms the latent variables into $\bm{Y}$.
This decomposition allows considering the correlation structure of treatment effects. 

\quad 
If the regression coefficient of matrix $\bm{\varGamma}_{\rm{(true)}}$ for the true treatment effects is full rank, for estimation $\bm{\varGamma}^{\dagger\top}\bm{x}_{i\rm{(r.v.)}}$ obtained through the weighting method based on ordinary multivariate regression, where ${\varGamma}^{\dagger}$ is estimated treatment effects based on weighting method for multiple outcomes with linear function (Yuki et al. 2023), 
the following equality holds:
\begin{equation}
\label{weight_propa}
    E[{Y}_{i}^{(1)}-{Y}_{i}^{(-1)}]=\bm{\varGamma}^{\dagger\top}\bm{x}_{i\rm{(r.v.)}}.
\end{equation}
However, if $\bm{\varGamma}_{\rm{(true)}}$ is not full rank, the estimation of the weighting method based on an ordinary multivariate regression has not been discussed as to whether it satisfies Eq.(\ref{weight_propa}).
Such a situation occurs when the true treatment effects have correlations.
Therefore, we propose a method to ensure that 
\begin{equation}
\label{weight_pro}
    E[{Y}_{i}^{(1)}-{Y}_{i}^{(-1)}]=\bm{\varGamma}^{\top}\bm{x}_{i\rm{(r.v.)}}
\end{equation}
holds even when $\bm{\varGamma}_{\rm{(true)}}$ is not full rank. To achieve this goal, we introduce the framework of the weighting approach for reduced-rank regression into the proposed method. 
Section 3 detail that the estimation of treatment effects in this framework satisfy Eq.(\ref{weight_pro}) in some situations.

\quad When the weighting method is applied to data containing outliers, as observed by Li et al.(2023), the estimation accuracy of treatment effects deteriorates. Therefore, we adopt the framework of robust reduced-rank regression proposed by She and Chen (2017) to handle outliers in the outcomes.
The objective function of robust reduced-rank regression can be formulated within the framework of regularized regression using parameters that adjust for outliers. It has been demonstrated in Theorem 2.2 of She and Chen (2017) that this objective function is equivalent to M-estimation. Consequently, robust estimation against outliers becomes feasible. Therefore, we focus on the framework and construct our objective function because this approach is easy to implement.

\quad From these things,
the model of the WMCMR4 is:
\begin{equation}
\label{model_vec}
Y_{i}=b(\bm{x}_{i\rm{(r.v.)}})+\frac{T_{i}}{2}\bm{V}_{\rm{(true)}}\bm{W}_{\rm{(true)}}^{\top}\bm{x}_{i\rm{(r.v.)}}+\bm{c}_{i\rm{(true)}}+\epsilon_{i} \quad \quad
 s.t.\quad\bm{V}^{\top}\bm{V}=\bm{I},
\end{equation}
where $\bm{c}_{i\rm{(true)}}\in\mathbb{R}^{q}$ are the true outlying effects on ${Y}_{i\rm{(r.v.)}}$ for subject $i$.

\quad Given $\bm{y}_{i\rm{(o.b)}}$, ${T}_{i\rm{(o.b.)}}$ and $\bm{x}_{i\rm{(o.b)}}$, the optimization problem of the WMCMR4 is defined as follows:
\begin{equation}
\label{mokuteki_vec}
\min_{\bm{W},\bm{V},\bm{c}_{i}}\sum_{i=1}^{n}\left|a_{i}\left({Y}_{i\rm{(o.b.)}}-{T}_{i\rm{(o.b.)}}\bm{V}\bm{W}^{\top}\bm{x}_{i\rm{(o.b.)}}/2-\bm{c}_{i}\right)\right|_{2}^{2}+\phi\sum_{i=1}^{n}|\bm{c}_{i}|_{2}+\lambda\sum_{k=1}^{p}|\bm{w}_{k}|_{2} \; \,
s.t. \; \, \bm{V}^{\top}\bm{V}=\bm{I},
\end{equation}
where $\phi(>0)$ and $\lambda(>0)$ are regularization parameters and $\bm{W}=(\bm{w}_{1}, \bm{w}_{2}, \ldots\bm{w}_{p})^{\top}$.
$|\bm{m}|_2$ is defined as $\ell2-$norm of an arbitrary vector $\bm{M}\in\mathbb{R}^{a}$.
Furthermore, ${a}_{i}$ is denoted as using the propensity score for $i$-th subject $\pi(\bm{x}_{i\rm{(o.b.)}})$ as follows:
\begin{equation*}
{a}_{i}=\frac{1}{\sqrt{{T_{i(\rm{(o.b.)})}\pi({\bm{x}_{i\rm{(o.b.)}}})+(1-T_{i(\rm{(o.b.)}})/2}}}.
\end{equation*}
This weight with a propensity score (Rosenbaum and Rubin 1983) is incorporated for each subject and enables application to data for both RCTs and observational studies.
This is the same as the weights in the weighting method.
The first term of Eq.(\ref{mokuteki_vec}) is a special case of the weighting approach.
In fact, 
\begin{equation*}
E_{{Y}_{i},T_{i}}\left[\left.\frac{\left\|{Y}_{i}-T_{i}\bm{V}\bm{W}^{\top}\bm{x}_{i\rm{(r.v.)}}/2\right\|_{2}^{2}} {T_{i}\pi({\bm{x}_{i\rm{(r.v.)}}})+(1-T_{i})/2}\right|\bm{x}_{i\rm{(r.v.)}}=\bm{
x}_{i\rm{(o.b.)}}\right]
\end{equation*}
corresponds to the first term of Eq.(\ref{mokuteki_vec}).

\quad In the first term of the objective function, by decomposing the regression coefficient matrix for the treatment effects, we can restrict the rank of the regression coefficient matrix for the treatment effects to at most $r$.
Specifically, $\bm{V}\bm{W}^{\top}\bm{x}_{i\rm{(o.b.)}}$ represents the treatment effect on the low-rank space.
Additionally, in the second term, we introduce a regularization term on $\bm{C}$. When the residuals corresponding to subject $i$ are small, this regularization encourages $\bm{c}_{i}$ to be sparse.
When the residuals corresponding to subject $i$ are large, the observations for subject $i$ are considered as outliers and this regularization encourages $\bm{c}_{i}$ to become non-zero.
This prevents excessive influence from the observations of subject $i$, identified as outliers.
Furthermore, in the third term of the objective function, we introduce a regularization term on $\bm{W}$ to induce the sparsity of rows corresponding to covariates unrelated to treatment effects for all outcomes. 
Consequently, it becomes easier to identify effective subgroups for treatment.

\quad Here, we express the model in Eq.(\ref{model_vec}) and the objective function in Eq.(\ref{mokuteki_vec}) in matrix form. Then, the model of the WMCMR4 is described as:
\begin{equation*}
\bm{Y}=B(\bm{X}_{\rm{(r.v.)}})+\frac{\bm{T}}{2}\bm{X}_{\rm{(r.v.)}}\bm{W}_{\rm{(true)}}\bm{V}_{\rm{(true)}}^{\top}+\bm{C}_{\rm{(true)}}+\bm{E} \quad \quad
 s.t.\quad\bm{V}^{\top}\bm{V}=\bm{I},
\end{equation*}
and the objective function of the WMCMR4 is:
\begin{equation}
\label{mokuteki}
\min_{\bm{W},\bm{V},\bm{C}}\left\|\bm{A}\left(\bm{Y}_{\rm{(o.b.)}}-\bm{T}_{\rm{(o.b.)}}\bm{X}_{\rm{(o.b.)}}\bm{W}\bm{V}^{\top}/2-\bm{C}\right)\right\|_{F}^{2}+\phi\|\bm{C}\|_{2.1}+\lambda\|\bm{W}\|_{2.1} \quad 
 s.t.\quad\bm{V}^{\top}\bm{V}=\bm{I},
\end{equation}
where $\bm{C}=(\bm{c}_{1}, \bm{c}_{2}, \ldots \bm{c}_{n})^{\top}\in\mathbb{R}^{n\times p}$ are the estimating outlying effects on $\bm{Y}_{\rm{(o.b.)}}$ , 
$B(\bm{X}_{(r.v.)})=(b_{1(\bm{x}_{(r.v.)})}, b_{2(\bm{x}_{(r.v.)})}, \ldots b_{n(\bm{x}_{(r.v.)})}): \mathbb{R}^{n\times p}\rightarrow\mathbb{R}^{n\times q}$ is a function of the true main effects, and
$\|\bm{M}\|_F$ and $\|\bm{M}\|_{2.1}$ are defined as the Frobenius norm and $\ell2.1-$norm of matrix $\bm{M}\in\mathbb{R}^{a\times b}$.
Furthermore, $\bm{A}={\rm{diag}}({a}_{1},{a}_{2},\ldots,{a}_{n})$ is a diagonal matrix representing the weights for each subject.

\subsection{Algorithm}
\quad  Here, we describe the algorithm for estimating $\bm{C}$, $\bm{W}$, and $\bm{V}$ in the WMCMR4. These parameters are estimated by the alternating least squares method. The following updated formula is used to estimate each parameter. We denote the estimationw of $\bm{C}$, $\bm{W}$, and $\bm{V}$ in the $t$-th step as $\bm{C}^{(t)}$, $\bm{W}^{(t)}$, and $\bm{V}^{(t)}$, respectively. We treat $\bm{X}$, $\bm{Y}$, and $\bm{T}$ as observed variables. 
\begin{description}
    \item[Updated formula for $\bm{C}$:] 
Given $\bm{W}^{(t-1)}$ and $\bm{V}^{(t-1)}$, 
the updated formula of $\hat{C}^{(t)}$ by the proximal gradient descent is derived as:

\begin{align*}
\hat{\bm{c}}^{(t)}_{i}\leftarrow\frac{1}{a_{i}^{2}}\left(1-\frac{\lambda}{2\|\bm{a}_{i}^{\top}\bm{R}^{\prime(t)}\|_{2}}\right)_{+}\bm{R}^{\prime(t)\top}\bm{a}_{i},
\end{align*}
where $\hat{\bm{c}}_{i}^{(t)}\in\mathbb{R}^{q}$ and $\bm{a}_{i}\in\mathbb{R}^{n}$ are the $i$-th row of $\hat{\bm{C}}^{(t)}$ and $\bm{A}$, respectively.
Furthermore, $\bm{R}^{\prime(t)}$ is expressed as:
\begin{align*}
\bm{R}^{\prime(t)}=-\bm{A}_{(-i)}\hat{\bm{C}}_{-i}^{(t)}+\bm{AY}-\bm{AZW}^{(t-1)}\bm{V}^{(t-1)\top},
\end{align*}
where $\bm{J}_{(-i)}$ and $\bm{J}_{-i}$ are the matrices obtained by removing the $\ell$-th column and $i$-th row of $\bm{J}$, respectively.
$\lambda(>0)$ is the regularization parameter and $(x)_{+}$ is $\max\{0,x\}$. We used the subgradient method, as per Chen and Huang (2012), for the updating formula of the sparse reduced rank regression.

\item[Updated formula for $\bm{W}$:] 
Given $\bm{C}^{(t)}$ and $\bm{V}^{(t-1)}$ 
, the updated formula of $\hat{W}^{(t)}$ by the proximal gradient descent is derived as:

\begin{align*}
\hat{\bm{w}}^{(t)}_{i}\leftarrow\frac{1}{\bm{f}_{(i)}^{(t)\top}\bm{f}^{(t)}_{(i)}}\left(1-\frac{\phi}{2\|\bm{f}_{(i)}^{(t)\top}\bm{R}^{\prime\prime(t)}\|_{2}}\right)_{+}\bm{R}^{\prime\prime(t)\top}\bm{f}_{(i)}^{(t)},
\end{align*}
where $\hat{\bm{w}}^{(t)}_{i}\in\mathbb{R}^{r}$ is the $i$-th row of $\hat{\bm{W}}^{(t)}$.
Furthermore, $\bm{R}^{\prime(t)}$ is expressed as:
\begin{align*}
\bm{R}^{(t)\prime\prime}=-\bm{AZ}_{(-i)}\hat{\bm{W}}^{(t)}_{-i}+(\bm{AY}-\bm{AC}^{(t)})\bm{V}^{(t-1)},
\end{align*}
and 
$\phi$ is the regularization parameter.
\item[Updated formula for $\bm{V}$:] Given $\bm{C}^{(t)}$ and $\bm{W}^{(t)}$, the updated formula for $\hat{W}^{(t)}$ by the Eckart--Young theorem (Eckart and Young 1936) is derived as:
:
\begin{align*}
\hat{\bm{V}}^{(t)}\leftarrow\bm{U}^{(t)}\bm{S}^{(t)\top},
\end{align*}
where $\bm{U}^{(t)}\bm{D}^{(t)}\bm{S}^{(t)\top}$ is the singular value decomposition of $\bm{\hat{W}}^{(t)\top}\bm{G}^{\top}\bm{F}^{(t)}$.
Here, $\bm{U}^{(t)}\in\mathbb{R}^{r\times k}$ and $\bm{S}^{(t)}\in\mathbb{R}^{q\times k}$ are left-
and right-singular vectors, respectively. $\bm{D}^{(t)}\in\mathbb{R}^{k\times k}= {\rm{diag}}(d_{1}, d_{2}, \ldots, d_{k})$ is the square diagonal matrix.
\end{description}

\quad The pseudo-code for the proposed method WMCMR4 is presented in Algorithm 1, where the objective function of WMCMR4 is denoted as ${L}$.
\begin{figure}
\begin{algorithm}[H]
    \caption{WMCMR4}
    \label{alg1}
    \begin{algorithmic}[1]    
    \REQUIRE$\bm{X}$,$\bm{Y}$,$\bm{T}$,$\bm{A}$,$\lambda$,$\phi$
    \ENSURE $\bm{W},\bm{V},\bm{C}$
    \STATE $\varepsilon>0$
    \STATE $t\leftarrow 1$
    \STATE $\bm{Z}\leftarrow \bm{T}\bm{X}/2$
    \STATE Set initial value for $\bm{C}^{(0)}$,$\bm{W}^{(0)}$,$\bm{V}^{(0)}$ 
    \WHILE{$L(\bm{W}^{(t)}$,$\bm{V}^{(t)}$,$\bm{C}^{(t)}|\bm{Z}$,$\bm{Y}$,$\bm{A}$,$\lambda$,$\phi)-L(\bm{W}^{(t-1)}$,$\bm{V}^{(t-1)}$,$\bm{C}^{(t-1)}|\bm{Z}$,$\bm{Y}$,$\bm{A}$,$\lambda$,$\phi)<\varepsilon$}
    \STATE$\varepsilon^{\prime}>0$
    \STATE$t^{*}\leftarrow 1$
    \WHILE{$\bm{C}^{(t^{*})}-\bm{C}^{(t^{*}-1)}<\varepsilon^{\prime}$}
    \FOR{$i=1$ to $n$}
    \STATE  $\bm{R}^{\prime{(t^{*}-1)}}\leftarrow-\bm{A}_{(-i)}{C}^{(t^{*}-1)}_{-i}+(\bm{AY}-\bm{AZW}^{(t-1)}\bm{V}^{(t-1)\top})$
    \STATE ${\bm{c}}_{i}^{(t^{*})}\leftarrow\frac{1}{a_{i}^{2}}\left(1-\frac{\lambda}{2\|\bm{a}_{i}^{\top}\bm{R}^{\prime{(t^{*}-1)}}\|_{2}}\right)_{+}\bm{R}^{\prime{(t^{*}-1)}\top}\bm{a}_{i}$
    \ENDFOR
    \STATE$t^{*}\leftarrow t^{*}+1$
    \ENDWHILE
    \STATE$\bm{C}^{(t)}\leftarrow \bm{C}^{(t^{*})}$
    \STATE $\bm{G}\leftarrow\bm{AZ}$ 
    \STATE $\bm{F}^{(t)}\leftarrow\bm{AY}-\bm{AC}^{(t)}$

    \STATE$\varepsilon^{\prime\prime}>0$
    \STATE$t^{**}\leftarrow 1$

    \WHILE{$\bm{W}^{(t^{**})}-\bm{W}^{(t^{**}-1)}<\varepsilon^{\prime\prime}$}
    
    \FOR{$i=1$ to $p$}
    \STATE $\bm{R}^{\prime\prime{(t^{**}-1)}}\leftarrow-\bm{G}_{(-i)}\bm{W}_{-i}^{(t^{**}-1)}+\bm{F}^{(t)}\bm{V}^{(t-1)\top}$
    \STATE $\bm{w}^{(t^{*})}_{i}\leftarrow\frac{1}{\bm{f}_{(i)}^{(t)\top}\bm{f}^{(t)}_{(i)}}\left(1-\frac{\phi}{2\|\bm{f}_{(i)}^{(t)\top}\bm{R}^{\prime\prime(t^{**}-1)}\|_{2}}\right)_{+}\bm{R}^{\prime\prime(t^{**}-1)\top}\bm{f}_{(i)}^{(t)}$
    \ENDFOR
    \STATE$t^{**}\leftarrow t^{**}+1$
    \ENDWHILE
    \STATE$\bm{W}^{(t)}\leftarrow \bm{W}^{(t^{**})}$

    \STATE  $\bm{U}^{(t)}\bm{D}^{(t)}\bm{S}^{(t)}$ $\leftarrow$ singular value decomposition of $\bm{W}^{(t)\top}\bm{G}^{\top}\bm{F}^{(t)}$
     \STATE $\bm{V}^{(t)}\leftarrow\bm{U}^{(t)}\bm{S}^{(t)\top}$
     \STATE $t \leftarrow t+1$
    \ENDWHILE
      \STATE$\bm{W} \leftarrow \bm{W}^{(t)}$
      \STATE$\bm{V} \leftarrow \bm{V}^{(t)}$
    \STATE$\bm{C} \leftarrow \bm{C}^{(t)}$
    \RETURN $\bm{W},\bm{V},\bm{C}$
    \end{algorithmic}
\end{algorithm}
\end{figure}
\section{Properties of the objective function}
\quad Here, we describe the relationship between the estimator obtained from WMCMR4 and the treatment effects. The weighting method showed that the estimator that minimizes the expected value for the objective function indicates the true treatment effect. Furthermore, Yuki et al. (2023), based on Chen et al. (2017), state that the true treatment effects can be obtained even when the weighting method is extended to handle multiple outcomes. However, in this method, no properties are known for the estimator when rank constraints are imposed on the regression coefficient matrix for the treatment effects. Therefore, we investigate the properties of the estimator that minimizes the expected value of the objective function of a weighting method based on reduced rank regression.

\subsection{Properties of WMCMR4}
\quad We consider the situation when there is a correlation in the treatment effects and a rank constraint is imposed on the regression coefficient matrix representing the treatment effects. Therefore, we use the framework of regularized rank regression for estimating the treatment effects.
Here, we prove that, under the assumption of estimating treatment effects at the correct rank, the expected value of the estimator for the weighting method based on reduced rank regression is equal to the CATE.

\quad Yuki et al. (2023) stated that treatment effects can be estimated when the objective function of the weighting method is based on the multivariate regression framework for treatment effects with full rank. 
However, the properties of the estimator of weighting method based on reduced rank regression have not been discussed in the literature. 
Under the reduced rank regression framework, this cannot be applied to the property because the optimization problem with the constraint is not the problem of the convex function.
Therefore, we describe the properties of the estimator obtained from minimizing the objective function, which is a weighting method based on a reduced rank regression.
Specifically, we show that the expected value of the estimator is the CATE when the potential outcomes assigned to the treatment or control groups are in a low-dimensional space.
Furthermore, we prove that the expected value of the estimator is equivalent to the true treatment effects, that is, \ref{weight_pro} holds if the true treatment effects exist on the low dimensional space.
First, we introduce models in which Eqs. (\ref{senzaiout_model1}) and (\ref{senzaiout_model-1}) are expressed in matrix notation to simplify the proof:
\begin{equation*}
\bm{Y}^{(1)}={B}(\bm{X}_{{\rm{(r.v.)}}})+\frac{\bm{1}}{2}\bm{X}_{{\rm{(r.v.)}}}\bm{\varGamma}_{\rm{(true)}}+\bm{E}, \\
\end{equation*}
\begin{equation*}
\bm{Y}^{(-1)}={B}(\bm{X}_{{\rm{(r.v.)}}})-\frac{\bm{1}}{2}\bm{X}_{{\rm{(r.v.)}}}\bm{\varGamma}_{\rm{(true)}}+\bm{E}.
\end{equation*}

\quad Lemmas \ref{hodai4_1} and \ref{hodai4_6} are described as preparation for explaining the properties of the WMCMR4.
\begin{lemma}
\label{hodai4_1}
\quad Let $Y_{i}\in\mathbb{R}^{q}$ be the vector of random variables for the outcomes, $T_{i}\in\{-1,1\}$ be the random variable for the treatment assignment, and $\bm{x}_{i\rm{(r.v.)}}\in\mathbb{R}^{p}$ be the random variable vector for the covariates. 
$\bm{y}_{i\rm{(o.b.)}}$, $t_{i}$, and $\bm{x}_{i\rm{(r.v.)}}$ correspond to their respective observed values.
Then,
\begin{equation*}
\begin{split}
&E_{{Y}_{i},T_{i}}\left[\left.\frac{\left\|{Y}_{i}-T_{i}\bm{\varGamma}^{\top}\bm{x}_{i\rm{(r.v.)}}/2\right\|_{2}^{2}} {T_{i}\pi({\bm{x}_{i\rm{(r.v.)}}})+(1-T_{i})/2}\right|\bm{x}_{i\rm{(r.v.)}}=\bm{
x}_{i\rm{(o.b.)}}\right]\\
=&E_{{Y}_{i}}\left.\left[\left\|{Y}_{i}-\bm{\varGamma}^{\top}\bm{x}_{i\rm{(r.v.)}}/2\right\|_{2}^{2}\right|T_{i}=1,\bm{x}_{i\rm{(r.v.)}}=\bm{x}_{i\rm{(o.b.)}}\right]\\&+E_{{Y}_{i}}\left.\left[\left\|{Y}_{i}+\bm{\varGamma}^{\top}\bm{x}_{i\rm{(r.v.)}}/2\right\|_{2}^{2}\right|T_{i}=-1,\bm{x}_{i\rm{(r.v.)}}=\bm{x}_{i\rm{(o.b.)}}\right] 
\end{split}
\end{equation*}
\end{lemma}

\begin{lemma}
\label{hodai4_6}
Let $\bm{Y}\in\mathbb{R}^{n\times q}$ be the random variable matrix for the outcomes, $\bm{X}_{\rm{(r.v.)}}\in\mathbb{R}^{n\times p}$ the random variable matrix for the covariates, and $\bm{T}\in\mathbb{R}^{n\times n}$ the diagonal matrix representing the treatment assignment. 
$\bm{Y}^{(1)}\in\mathbb{R}^{n\times q}$ and $\bm{Y}^{(-1)}\in\mathbb{R}^{n\times q}$ are the potential outcomes matrices if all subjects are assigned to the treatment or control groups, respectively.
Then, the following equation holds:
\begin{align*}
&E_{\bm{Y}^{(1)},\bm{Y}^{(-1)}}\left[\|\bm{Y}-\bm{X}^{*}_{\rm{(r.v.)}}\bm{\varGamma}\|_{F}^{2}|\bm{T}=\bm{I},\bm{X}_{\rm{(r.v.)}}=\bm{X}_{\rm{(o.b.)}}\right]\\
&+E_{\bm{Y}^{(1)},\bm{Y}^{(-1)}}\left[\|\bm{Y}+\bm{X}^{*}_{\rm{(r.v.)}}\bm{\varGamma}\|_{F}^{2}|\bm{T}=-\bm{I},\bm{X}_{\rm{(r.v.)}}=\bm{X}_{\rm{(o.b.)}}\right]\\
=&E_{\bm{Y}^{(1)},\bm{Y}^{(-1)}}\left.\left[\left\|\frac{1}{\sqrt{2}}(\bm{Y}^{(1)}-\bm{Y}^{(-1)})-\sqrt{2}\bm{X}^{*}_{\rm{(r.v.)}}\bm{\varGamma}\right\|_{F}^{2}\right|\bm{X}_{\rm{(r.v.)}}=\bm{X}_{\rm{(o.b.)}}\right]\\
&+E_{\bm{Y}^{(1)},\bm{Y}^{(-1)}}\left[\left.{\rm{tr}}\left(\frac{1}{2}\bm{Y}^{(1)\top}\bm{Y}^{(1)}-\frac{3}{2}\bm{Y}^{(1)\top}\bm{Y}^{(-1)}+\frac{1}{2}\bm{Y}^{(-1)\top}\bm{Y}^{(-1)}\right)\right|\bm{X}_{\rm{(r.v.)}}=\bm{X}_{\rm{(o.b.)}}\right].
\end{align*}
\end{lemma}
\quad We show that the parameters that minimize the expected value of the weighting method based on reduced rank regression can express CATE on the low-rank space in the next theorem using Lemmas \ref{hodai4_1} and \ref{hodai4_6}.

\begin{theorem}
\label{theorem}
Let $\bm{\varGamma}\in\mathbb{R}^{p\times q}$ be the regression coefficient matrix for the treatment effects.
We assume that ${\rm{rank(\bm{\varGamma})}}=r\,(\leq \min(p,q))$ and $\bm{X}^{\top}_{{\rm{(o.b.)}}}\bm{X}_{{\rm{(o.b.)}}}$ is regular. 
Here, we also define $\bm{L}_{w}(\bm{\varGamma}|\bm{X}_{\rm{(r.v.)}}=\bm{X}_{\rm{(o.b.)}})$ as Eq. ({\ref{l_w}}):
\begin{equation}
\label{l_w}
\bm{L}_{w}\left(\bm{\varGamma}\left|\bm{X}_{\rm{(r.v.)}}=\bm{X}_{{\rm{(o.b.)}}}\right)\right.= E_{\bm{Y},\bm{T}}\left[\sum_{i=1}^{n}\frac{\|Y_{i}-T_{i}\bm{\varGamma}^{\top}\bm{x}_{i {\rm{(r.v.)}}}/2\|_{F}^{2}} {T_{i}\pi({\bm{x}_{i{\rm{(r.v.)}}}})+(1-T_{i})/2}\Big{|}\bm{X}_{{\rm{(r.v.)}}}=\bm{
 X}_{{\rm{(o.b.)}}}\right].
\end{equation}
The estimator of $\bm{\varGamma}$ that minimizes $\bm{L}_{w}(\bm{\varGamma}|\bm{X}_{{\rm{{\rm{(r.v.)}}}}}=\bm{X}_{\rm{(o.b.)}})$ is denoted as $\hat{\bm{\varGamma}}$. 
For $\hat{\bm{\varGamma}}$, there exist $\hat{\bm{W}}\in\mathbb{R}^{p\times r}$ and $\hat{\bm{V}}
\in\mathbb{R}^{q\times r}$ that satisfy $\hat{\bm{\varGamma}}=\hat{\bm{W}}\hat{\bm{V}}^{\top}$, where $\hat{\bm{V}}$ is the column orthogonal matrix.
In this situation, the following equality holds:
\begin{align*}
E_{\bm{Y}^{(1)},\bm{Y}^{(-1)}}\left[\bm{Y}^{(1)}\bm{V}^{*}\bm{V}^{*\top}-\bm{Y}^{(-1)}\bm{V}^{*}\bm{V}^{*\top}|\bm{X}_{{\rm{(r.v.)}}}=\bm{X}_{{\rm{(o.b.)}}}\right]
&=\bm{X}_{{\rm{(o.b.)}}}\bm{W}^{*}\bm{V}^{*\top}.
\end{align*}
where $\hat{\bm{W}}^{*}=E_{\bm{Y}^{(1)},\bm{Y}^{(-1)}}\left[\hat{\bm{W}}\right]$ and $\hat{\bm{V}}^{*}=E_{\bm{Y}^{(1)},\bm{Y}^{(-1)}}\left[\hat{\bm{V}}\right]$.
\end{theorem}
In the Theorem \ref{theorem}, we showed that the estimator expresses the expected value of the differences in potential outcomes for the treatment and control groups on a low-dimensional space.
As such, we introduce Corollary \ref{wp_meidai9} of this theorem.
\begin{corollary}
\label{wp_meidai9}
We assume that $\bm{\varGamma}_{\rm{(true)}}$ can be decomposed by $\bm{\varGamma}^{*}\in\mathbb{R}^{p\times r}$ and the orthogonal matrix $\bm{V}^{*}\in\mathbb{R}^{q\times r}$, that is, $\bm{\varGamma}_{\rm{(true)}}=\bm{\varGamma}^{*}\bm{V}^{*\top}$.
Then, the following equality holds:
\begin{align*}
E\left[\bm{Y}^{(1)}-\bm{Y}^{(-1)}|\bm{X}_{\rm{(r.v.)}}=\bm{X}_{\rm{(o.b.)}}\right]
=\bm{X}_{\rm{(o.b.)}}{\bm{W}}^{*}\bm{V}^{*\top}
\end{align*}
\end{corollary}
We showed that the expected value of the estimated parameter is the true treatment effect if $\mathrm{rank}(\bm{\varGamma})=\mathrm{rank}(\bm{\varGamma}_{(\mathrm{true})})$.
Eventually, we can use the estimator obtained by the weighting method based on the reduced rank regression to approximate the CATE if the treatment effect can be estimated with the true rank.

\section{Simulation Study}
To reveal the properties of the proposed method, we present here a numerical simulation.
In this simulation, the results are compared to the control methods for various situations, that is, including outliers and rank of true treatment effects.
\subsection{Simulation design}
\quad In the simulation study, we considered both RCTs and observational studies.
We referred to Tian et al. (2014), Chen et al. (2017), Tan et al. (2023) for the simulation design.
In the RCT, we assumed that the propensity score becomes 0.5.
The outcome for subject $i$ is generated as follows:
\begin{align*}
\bm{y}_{i\rm{(o.b.)}}=\bm{B}^{\top}\bm{x}_{i\rm{(o.b.)}}\odot\bm{B}^{\top}\bm{x}_{i\rm{(o.b.)}}+T_{i\rm{(o.b.)}}(\bm{\varGamma}^{\top}\bm{x}_{i\rm{(o.b.)}})/2+\bm{\varepsilon}_{i},
\end{align*}
where the $\odot$ is a Hadamard product and $\bm{x}_{i\rm{(o.b.)}}=(1,\bm{x}_{i\rm{(o.b.)}}^{*\top})^{\top}\in\mathbb{R}^{p+1}$ are the covariates for subject $i$, including the intercept. 
In this formula, the main effects are non-linear to show that the proposed method is not affected by the main effects. Assuming the presence of outliers in the outcomes for $\tau\%$ of the overall subjects, some outcomes are generated as follows: $\bm{y}_{i\rm{(o.b.)}}\sim U(15,20)$.

To accommodate outliers in the outcome, the outcome for subjects with $\tau\%$ is generated from $\bm{y}_{i\rm{(o.b.)}}\sim U(15,20)$.
We generate covariate $\bm{x}_{i\rm{(o.b.)}}^{*\top}$ for subject $i$ from a multivariate normal distribution $N(\bm{0}_{p},\Sigma_{x})$, where $\bm{0}_{a}\in\mathbb{R}^{a}$ is a vector with all elements zero and $\Sigma_x\in\mathbb{R}^{p\times p}$ is a symmetric matrix with the diagonal elements equal to $1$ and the off-diagonal elements equal to $g$.
The true regression coefficient matrix representing the main effect is $\bm{B}=(\bm{0}_{q},\bm{B}^{*\top})^{\top}\in\mathbb{R}^{(p+1)\times q}$ and $\bm{B}^{*}\in\mathbb{R}^{p\times q}$ is a matrix where all elements are $b$ in columns 3--10 and all other elements are 0.
The true regression coefficient matrix representing the treatment effects is $\bm{\varGamma}=(\bm{0}_{q},\bm{\varGamma}^{*\top})^{\top}\in\mathbb{R}^{(p+1)\times q}$. 
Error $\bm{e}_{i}\in\mathbb{R}^{q}$ for subject $i$ is generated to follow a multivariate normal distribution, $N(\bm{0}_q,\Sigma_e\in\mathbb{R}^{q\times q})$, where $\Sigma_e$ is a symmetric matrix with the diagonal elements equal to $2$ and the off-diagonal elements equal to $z$.

\quad In the RCT, we generated $T_{i}$ representing the treatment assignment for subject $i$ from a Bernoulli distribution, $B(1, 0.5)$.
However, in an observational study, we generated $T_{i}$ from the Bernoulli distribution $B(1, P_{i})$, where $P_{i}=1/\left(1+{\rm{exp}}(x_{i1}+x_{i2}+x_{i3}+x_{i4}+x_{i5})\right)$.
Subjects whose Bernoulli distribution produced $1$ were assigned to the treatment group and subjects whose Bernoulli distribution produced $0$ were transformed from $0$ to $-1$ and assigned to the control group.

\quad In all scenarios, the sample size for the training data $n$ is $300$, and $n^*$ is $1000$ for the test data.
The number of outcomes is set to $q=10$.
The number of iterations for each scenario is set to 100. 
We validated the six scenarios regarding the number of covariates and other factors.
\begin{enumerate}[(I)]
\item Number of covariates: $p=\{10,50\}$
\item Correlation of covariates: $g=\{0,1/3\}$
\item Proportion of outcome outliers: $\tau=\{0,5,10\}$
\item Elements of the main effects: $b=\{6^{-1/2},3^{-1/2}\}$
\item Rank and presence or absence of sparse elements in the true  regression coefficient matrix representing the treatment effects
\item Correlation of errors: $z=\{0,1/3\}$
\end{enumerate}
With respect to $($V$)$, we consider the following four scenarios:
\begin{enumerate}[(1)]
 \item Rank 1 and absence of sparse elements: $\bm{\varGamma}=\bm{u}_{1}\bm{v}_{1}^{\top}$, where each element of $\bm{u}_{1}\in\mathbb{R}^{p}$ and
 $\bm{v}_{1}\in\mathbb{R}^{q}$ is generated from a uniform distribution on the interval.
 \item  Rank 2 and absence of sparse elements: $\bm{\varGamma}=\bm{u}_{1}\bm{v}_{1}^{\top}+\bm{u}_{2}\bm{v}_{2}^{\top}$, where each element of $\bm{u}_{1}\in\mathbb{R}^{p}$, $\bm{u}_{2}\in\mathbb{R}^{p}$ and
 $\bm{v}_{1}\in\mathbb{R}^{q}$, $\bm{v}_{2}\in\mathbb{R}^{q}$ is generated from a uniform distribution on the interval.
 \item  Rank 1 and presence of sparse elements: 
 $\bm{\varGamma}=\bm{u}_{1}\bm{v}_{1}^{\top}$ with $\bm{u}_{1}\in\mathbb{R}^{p}=(\bm{1}_{4}^{\top},\bm{0}_{p-4}^{\top})^{\top}$ and $\bm{v}_{1}\in\mathbb{R}^{q}=
 (\bm{1}_{4}^{\top},\bm{0}_{q-4}^{\top})^{\top}$,
 $\bm{u}_{2}\in\mathbb{R}^{p}=\bm{0}_{p}$,$\bm{v}_{2}\in\mathbb{R}^{q}=\bm{0}_{q}$
 \item Rank 2 and presence of sparse elements: $\bm{\varGamma}=\bm{u}_{1}\bm{v}_{1}^{\top}+\bm{u}_{2}\bm{v}_{2}^{\top}$ with $\bm{u}_{1}=(\bm{1}_{4}^{\top},\bm{0}_{p-4}^{\top})^{\top}$,$\bm{u}_{2}\in\mathbb{R}^{p}=(\bm{0}_{2}^{\top},\bm{1}_{4}^{\top},\bm{0}_{p-6}^{\top})^{\top}$
 ,$\bm{v}_{1}=(\bm{1}_{4}^{\top},\bm{0}_{q-4}^{\top})^{\top}$ and 
 $\bm{v}_{2}\in\mathbb{R}^{q}=(\bm{0}_{2}^{\top},\bm{1}_{4}^{\top},\bm{0}_{q-6}^{\top})^{\top}$
\end{enumerate}
The regularization parameters ($\lambda$,$\phi$) of the proposed method
and the rank of $\bm{W}$ are determined by five-fold cross validation.
Eventually, the accuracy of the proposed method is evaluated based on the results of the 192 scenarios combining the situations from (I) to (VI).

\quad Subsequently, we describe the evaluation index for this simulation.
We employ MSE, bias, Spearman's rank correlation coefficient (spearman), and AUC as evaluation indices.
These indices are defined as follows:
\begin{align*}
  {\rm{MSE}}=\frac{1}{n^{*}q}\|\bm{X}^{\rm{test}}\hat{\bm{\varGamma}}-\bm{X}^{\rm{test}}{\bm{\varGamma}}\|_{F}^{2},\hspace{5mm}
  {\rm{bias}}=\frac{1}{n^{*}q}\left|\sum_{i=1}^{n^{*}}\sum_{j=1}^{q}(\bm{x}_{i}^{\rm{test}\top}\hat{\bm{\gamma}_{j}}-\bm{x}_{i}^{\rm{test}\top}{\bm{\gamma}_{j}})\right|,
\end{align*}
\begin{align*}
{\rm{Spearman}}=1-\frac{6\sum_{i=1}^{n^{*}}(\hat{d}_{i}-{d}_{i})^{2}}{n(n^{2}-1)},
\end{align*}
where $\bm{X}^{\rm{test}}=(\bm {x}^{\rm{test}}_{1},\bm{x}^{\rm{test}}_{2},\ldots,\bm {x}_{{n}^{*}}^{\rm{test}})^{\top}\in\mathbb{R}^{n^{*}\times(p+1)}$
are the covariates for the test data and $\bm{\hat{\varGamma}}=(\bm{\hat{\gamma}}_{1},\bm {\hat{\gamma}}_{2},\ldots,\bm{\hat{\gamma}}_{q})\in\mathbb{R}^{(p+1)\times q}$ 
is the regression coefficient matrix for the estimated treatment effects. 
Let $\hat{d}_i$ and $d_i$ denote the rank of the $i$-th subject when the $\hat{D}_i=\sum_{j=1}^{q}(\bm{x}_{i}^{\top\rm{test}}\hat{\bm{\gamma}_{j}})$ and ${D_{i}}=\sum_{j=1}^{q}(\bm{x}_{i}^{\top\rm{test}}{\bm{\gamma}_{j}})$ values for all subjects are arranged in descending order, respectively.
AUC assesses whether estimated treatment effects score $\hat{D_{i}}$ can discriminate ${\rm{sign}}{(D_{i})}$.
It is an index for assessing classification accuracy in machine learning and other applications using false negative and false positive rates, which are defined as follows:
\begin{align*}
    &{\rm{false\ negative}}
    \\
    &={\rm{card}}\left\{i\left|\sum_{j=1}^{q}(\bm{x}_{i}^{\rm{(test)\top}}\hat{\bm{\gamma}_{j}})\leq0\quad{\rm{and}}\quad \sum_{j=1}^{q}(\bm{x}_{i}^{\rm{(test)}\top}{\bm{\gamma}_{j}})\leq0\right. \right\}\big/
    {\rm{card}}\left\{i\left|\sum_{j=1}^{q}(\bm{x}_{i}^{\rm{(test)}\top}{\bm{\gamma}_{j})\leq 0}\right.\right\}
\end{align*}

\begin{align*} 
    &{\rm{false\ positive}}
    \\
    &={\rm{card}}\left\{i\left|\sum_{j=1}^{q}(\bm{x}_{i}^{\rm{(test)\top}}\hat{\bm{\gamma}_{j}})>0\quad{\rm{and}}\quad \sum_{j=1}^{q}(\bm{x}_{i}^{\rm{(test)}\top}{\bm{\gamma}_{j}})>0\right. \right\}\big/
    {\rm{card}}\left\{i\left|\sum_{j=1}^{q}(\bm{x}_{i}^{\rm{(test)}\top}{\bm{\gamma}_{j})> 0}\right.\right\},
\end{align*}
where ${\rm{card}}\{A\}$ is the number of elements of set $A$.

\quad We explain the methods employed for comparing the accuracy with that of the proposed method.
They consist of four methods: MCMRRR---a method that applies the reduced-rank regression (Izenman
1975) framework to MCM, MCM$\ell1$---a method that applies $\ell1.1-{\rm{norm}}$ to the MCM loss function(Li
et al. 2023), MCM---MCM with an extension of the multivariate regression framework (Tian et al.
2014), and Full---a method that estimates both main and treatment effects.
The objective function of each method is as follows:\\
(1)WMCMRRR
\begin{align*}
&\min_{\bm{W},\bm{V}}\left\|\bm{A}\bm{Y}-\bm{T}\bm{X}\bm{W}\bm{V}^{\top}/2\right\|_{F}^{2}+\lambda\|\bm{W}\|_{2.1}\quad\quad s.t.\quad\bm{V}^{\top}\bm{V}=\bm{I}
\end{align*}
(2)WMCM$\ell1$
\begin{align*}
&\min_{\bm{\varGamma}}\left\|\bm{A}\bm{Y}-\bm{T}\bm{X}\bm{\varGamma}/2\right\|_{1.1}+\lambda\|\bm{\varGamma}\|_{2.1}
\end{align*}
(3)WMCM
\begin{align*}
&\min_{\bm{\varGamma}}\left\|\bm{A}\bm{Y}-\bm{T}\bm{X}\bm{\varGamma}/2\right\|_{F}^{2}+\lambda\|\bm{\varGamma}\|_{2.1}
\end{align*}
(4)WFull
\begin{align*}
&\min_{\bm{B,\varGamma}}\left\|\bm{A}\bm{Y}-\bm{X}\bm{B}-\bm{T}\bm{X}\bm{\varGamma}/2\right\|_{F}^{2}+\lambda\|\bm{\varGamma}\|_{2.1},
\end{align*}
where $\|\bm{M}\|_F$ and $\|\bm{M}\|_{1.1}$ are defined as the Frobenius norm and $\ell1.1-$norm of matrix $\bm{M}\in\mathbb{R}^{a\times b}$, respectively.
In the RCT scenarios, the treatment effect is estimated with $\bm{A}$ set to identity matrix $\bm{I}$.
Therefore, the methods are called MCMRRR, MCM$\ell1$, MCM, and Full.

\subsection{Simulation results}
\quad Here, we discuss the results of MSE and Spearman's rank correlation coefficient when the number of covariates $p$ is set to $50$.
First, we focus on the results of MSE in the RCT scenarios.
In cases where outliers are absent, the proposed method demonstrates the smallest MSE in scenarios 1 and 2 for the regression coefficient matrices for treatment effects. 
However, there are no significant differences in the MSE across each method in scenarios 3 and 4.
In all scenarios, when 5\% of the outcomes for the overall subjects are outliers, our method consistently demonstrated the minimum MSE compared to other methods.
This suggests that the proposed method can accurately estimate treatment effects compared to other methods, even when the proportion of outliers is modest, without significantly influencing the presence or absence of correlation among covariates or the non-linearity of the main effect.
However, when 10\% of the outcomes for the overall subjects are outliers, the MSE of the proposed method increased and MCM$\ell1$ had better estimation accuracy in terms of MSE compared to our method. 
This indicates that the proposed method may fail to accurately estimate treatment effects for each subject compared to MCM$\ell1$ when the proportion of outliers is high.
For observational studies where outliers are absent, there are no significant differences in the MSE of each method in scenarios 1 and 2 in representing the regression coefficient matrices for treatment effects.
Moreover, in scenarios 3 and 4, our method tends to have a smaller MSE. 
However, when both covariates and errors exhibit correlation, the proposed method tends to increase the MSE.
When 5\% of the outcomes for the overall subjects are outliers, the proposed method exhibits the minimum MSE in scenarios 1 and 2 in representing the regression coefficient matrices for treatment effects. 
In scenarios 3 and 4, both the proposed method and MCM$\ell1$ demonstrate smaller MSE compared to the other three methods.
However, when 10\% of outliers are present among all subjects and there is a correlation among covariates, the MSE of our method is smaller than that of MCM$\ell1$.
Nonetheless, when there is a correlation among covariates, the MSE of MCM$\ell1$ increases, while the MSEs of MCM and Full tend to be smaller compared to those of the other methods.

\quad Next, we discuss the results of Spearman's rank correlation coefficient for RCTs. 
Here, the results of Spearman's rank correlation were set as $0$ when all elements of the coefficient vector were estimated as $0$.
When outliers are absent among all subjects, there are no significant differences in Spearman's rank correlation coefficient among the various methods in Scenarios 1 and 2 for regression coefficient matrices for the treatment effects.
In Scenarios 3 and 4 for the regression coefficient matrices for treatment effects, both the proposed method and $\ell$1MCM tend to have slightly higher Spearman's rank correlation coefficient values compared to the other methods.
When 5\% of outliers are present among all subjects, regardless of scenario, the Spearman's rank correlation coefficients for MCMRRR, MCM, and Full are observed to be smaller compared to the case where outliers are absent, 
while the rank correlation coefficients for the proposed method and MCM$\ell1$ are relatively larger. 
From these results, when 5\% of the overall subjects contain outliers, the proposed method can correctly identify subgroups compared to other methods. 
However, when 5\% of outliers are present among all subjects,
Spearman's rank correlation coefficient of the proposed method is small. 
Consequently, when the proportion of outliers is too high, the proposed method may not effectively identify subgroups.

\quad
We discuss the situation for observational studies as well.
When outliers are absent, in Scenarios 1 and 2 for the regression coefficient matrix for treatment effects, MCMRRR has a Spearman's rank correlation coefficients closer to 1. 
This indicates a strong positive correlation between estimated treatment effects and true treatment effects ranks, confirming that MCMRRR can most accurately identify subgroups. 
In Scenarios 3 and 4, all methods show a tendency for Spearman's rank correlation coefficients to be closer to 1.
Particularly, Spearman's rank correlation coefficients for the proposed method closely approach 1 and there exists a strong positive correlation between the estimated treatment effects and the true treatment effect ranks.
Therefore, the proposed method accurately identifies subgroups.
When 5\% of outliers are present among all subjects, it cannot be said that the MCMRRR, MCM, and Full exhibit positive correlations between the estimated treatment effects and the true treatment effect ranks.
However, both the proposed method and $ \ell1$MCM exhibit a linear correlation between the order of the estimated treatment effects and the order of the true treatment effects.
Particularly, in Scenarios 3 and 4, the rank correlation coefficients approach 1.
Hence, it can be inferred that both the proposed method and MCM$\ell1$ accurately identify subgroups.
When 10\% of outliers are present among all subjects, it is evident that the proposed method performs worse than MCM$\ell$1 based on Spearman's rank correlation coefficients.
Particularly in Scenarios 1 and 2, the accurate identification of subgroups cannot be achieved.
However, in Scenarios 3 and 4, the proposed method demonstrated a linear correlation between the order of the estimated treatment effects and the order of the true treatment effects. 
This suggests that the proposed method can reasonably identify subgroups.
The results when the number of variables, $p$, is set to 50 and the bias and AUC results are described in the Appendix.

\quad As a general trend, when the proportion of outliers in the outcome is high, the proposed method tends to provide inadequate, inaccurate estimations. 
This can be attributed to the fact that, as the proportion of outliers increases, there is a tendency for cross-validation to select ranks larger than the true ranks, resulting in a decrease in the estimation accuracy of the proposed method.
Moreover, the presence of correlations among covariates may lead to a decrease in the accuracy of treatment effect estimations, depending on the scenario.
This is also attributed to the errors in the selection of ranks during estimation.
Therefore, it is necessary to correctly establish ranks when estimating treatment effects using the proposed method.
\begin{figure}
\includegraphics[keepaspectratio, scale=0.9]{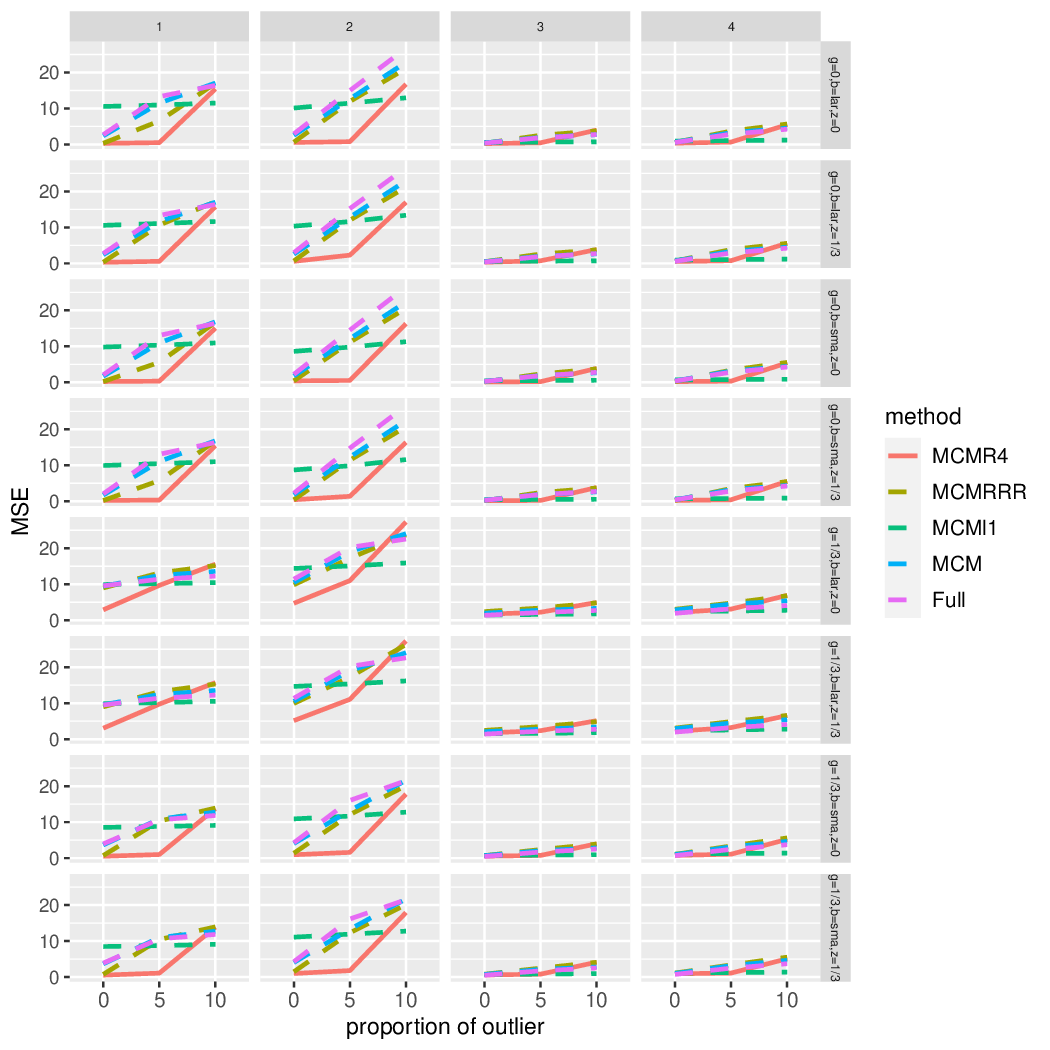}
    \label{sim_mse1}
      \centering{
    \caption{\small{Plot of MSE for RCT}}
    }
    \small{The horizontal axis of each plot indicates the proportion of outliers present in the outcome and the vertical axis represents the MSE values. The numbers on the horizontal axis of Figure \ref{sim_mse1} represent the scenarios for the regression coefficient matrix for treatment effects. The vertical axes, labeled g, b, and z, denote the value of covariate correlations, main effects, and error correlation, respectively. When b=sma, it represents $b=6^{(-1/2)}$ and, if b=lar, it represents $b=3^{(-1/2)}$.}
\end{figure}

\begin{figure}
    \includegraphics[keepaspectratio, scale=0.9]
    {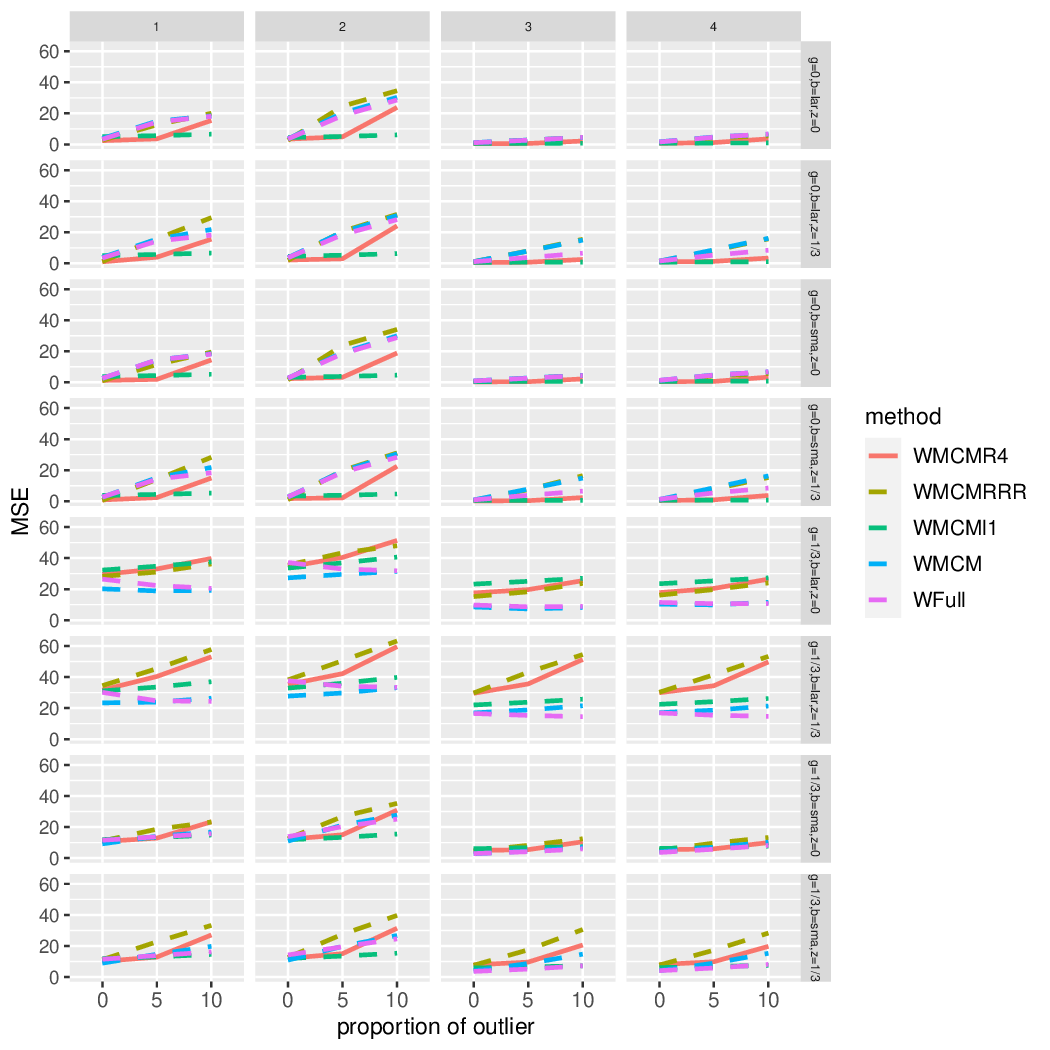}
      \centering
    \caption{\small{Plot of MSE for observational studies}}
    \label{sim_mse2}
\end{figure}

\begin{figure}
    \includegraphics[keepaspectratio, scale=0.9]{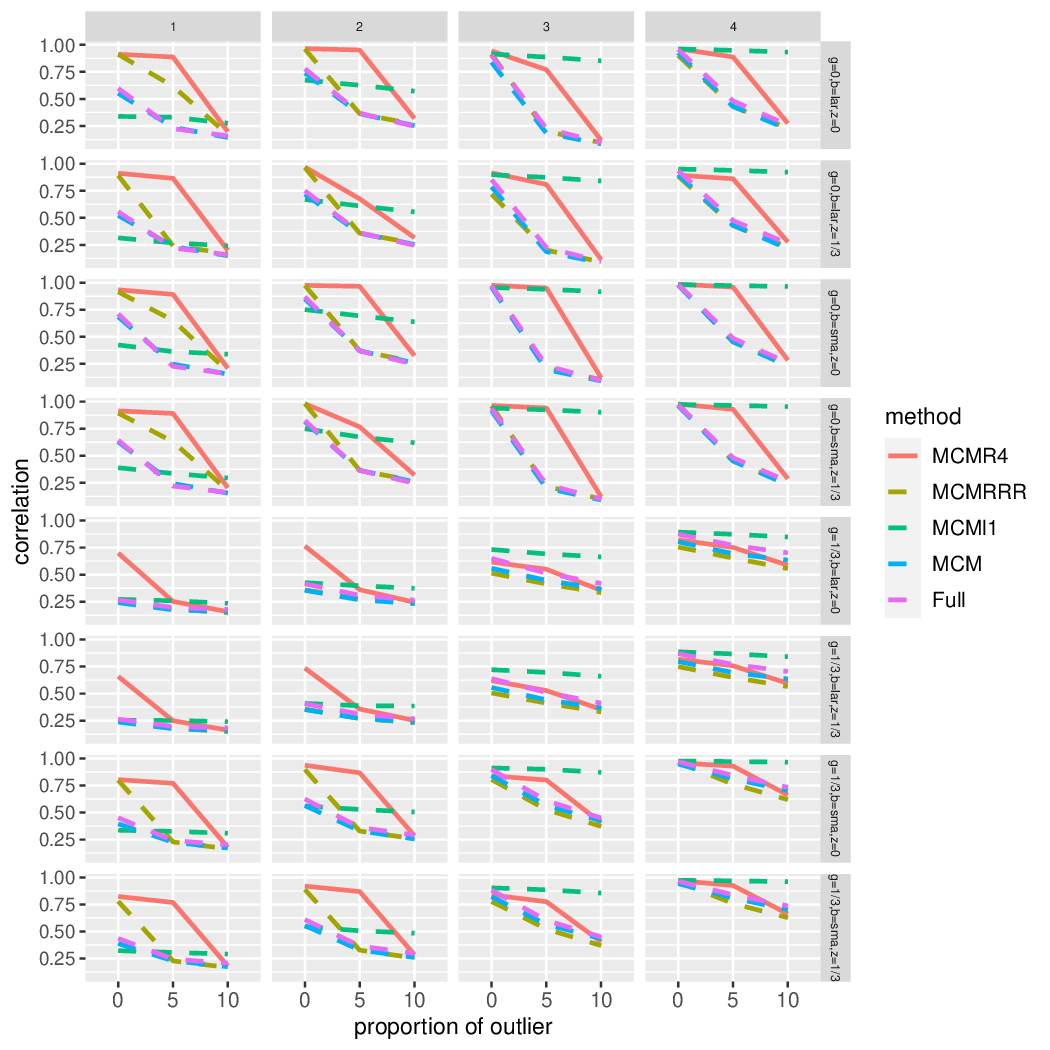}
    \centering
    \caption{\small{Plot of rank correlation for RCT}}
    \label{sim_rank correlation3}
\end{figure}

\begin{figure}
    \includegraphics[keepaspectratio, scale=0.9]{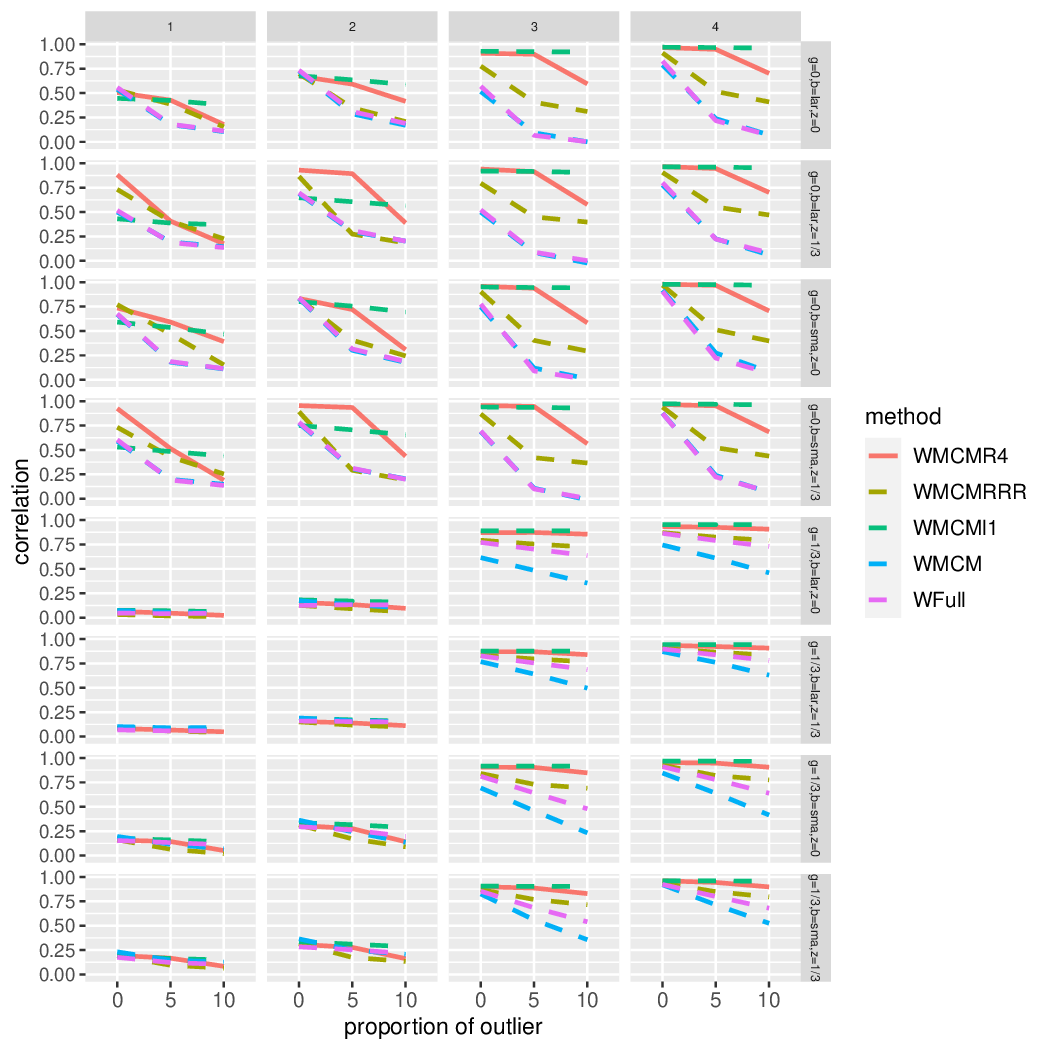}
    \caption{\small{Plot of rank correlation for observational studies}}
    \label{sim_rank correlation4}
\end{figure}

\section{Real Data Application}
\quad Here, we apply both the proposed method and comparison methods to real data. We demonstrate that the proposed method can more accurately identify subgroups compared to the comparison methods. Additionally, we provide examples of the subgroups identified by the proposed method.
We explain the dataset in Section \ref{reldata} and describe the methods and results in Section \ref{Interpretation}.

\subsection{Real Data}
\label{reldata}
\quad ACTG175 is a dataset from a RCT focused on adults infected with Human Immunodeficiency Virus Type 1 (HIV-1) with CD4 cell counts ranging from 200 to 500 cells per cubic millimeter.
This dataset can be used to investigate the differences in the effects of drugs on the counts of CD4 T and CD8 T cells across four groups (1: zidovudine, 
2: zidovudine and didanosine, 3: zidovudine and zalcitabine, 4: didanosine) of subjects. 
Here, we treat the subjects who take combination therapy of zidovudine and zalcitabine ($n=524$) as the treatment group and the subjects who take only zidovudine ($n=532$) as the control group.
ACTG175 is included in the ``speff2trial'' package in R (Juraska and Juraska 2022).

\begin{figure}
    \includegraphics[keepaspectratio, scale=0.4]{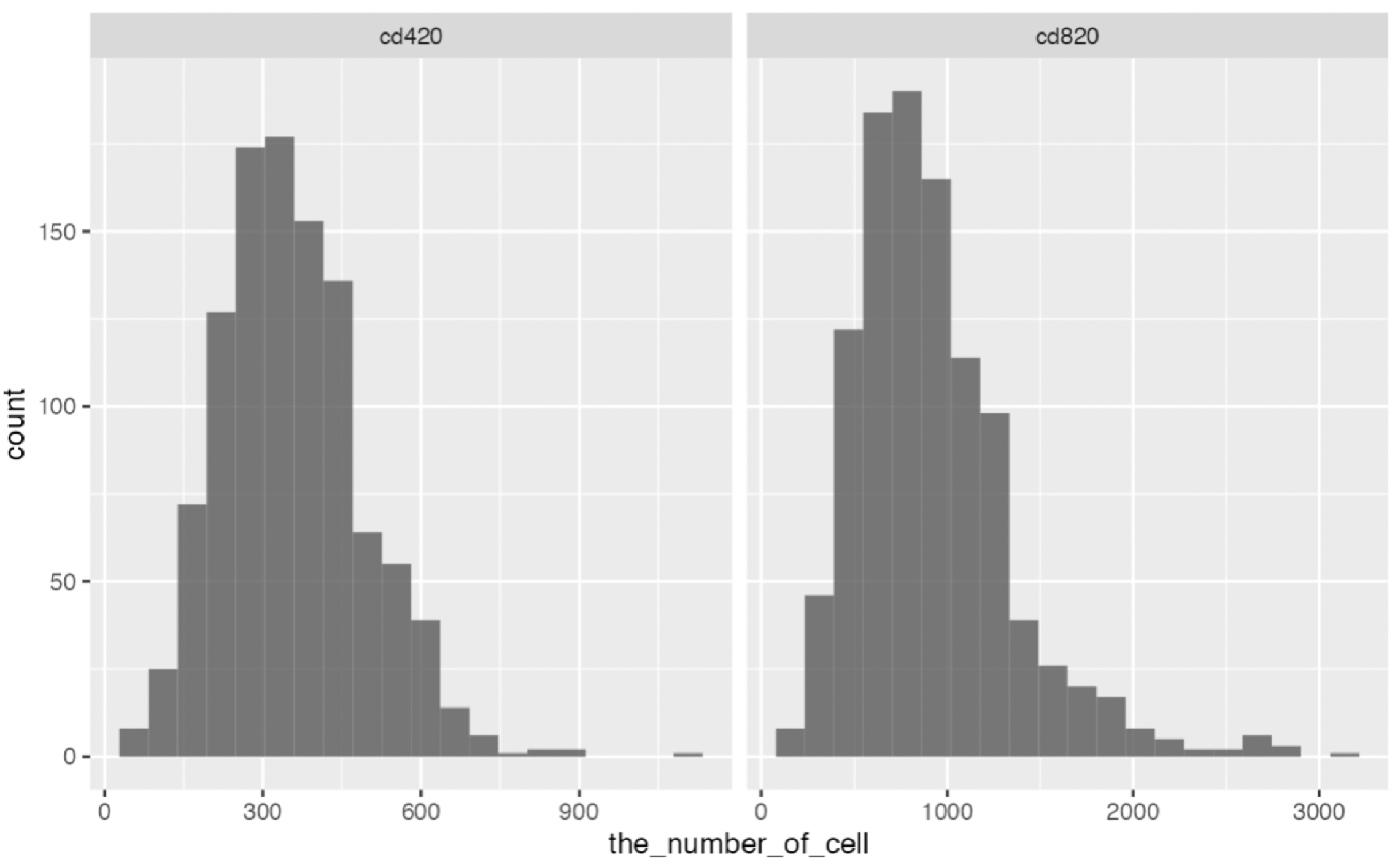}
     \caption{histgram of CD4T and CD8T cell count}
     \label{ACTG_histgram}
\end{figure}

\quad We take two variables outcomes: The CD4 T cell count at 20 $\pm$ 5 weeks (cd420) and the CD8 T cell count at 20 $\pm$ 5 weeks (cd820).
As illustrated in Figure \ref{ACTG_histgram}, both the distributions of CD4 T and CD8 T cell counts are positively skewed.
Therefore, both outcomes may contain outliers.
We set 14 covariates: age in years at baseline (age); 
weight in kg at baseline (wtkg); 
hemophilia: 0=no, 1=yes (hemo); homosexual activity:
0=no, 1=yes (homo); 
Karnofsky score(on a scale of 0--100) (karnof); 
CD4 T cell count at baseline (cd40); 
CD8 T cell count at baseline (cd80); 
zidovudine use in the 30 days prior to treatment initiation: 0=no, 1=yes (z30); 
race: 0=white, 1=non-white (race); 
history of intravenous drug use: 0=no, 1=yes (drugs);
gender: 0=female, 1=male (gender); 
antiretroviral history: 0=naive, 1=experienced (str2);
symptomatic indicator: 0=asymptomatic, 1=symptomatic (symp);
non-zidovudine antiretroviral therapy prior to initiation of study treatment: 0=no, 1=yes (oprior).

\subsection{Interpretation of regression coefficients for the proposed method}
\label{Interpretation}

\begin{figure}
\centering
\includegraphics[width=14.5cm]{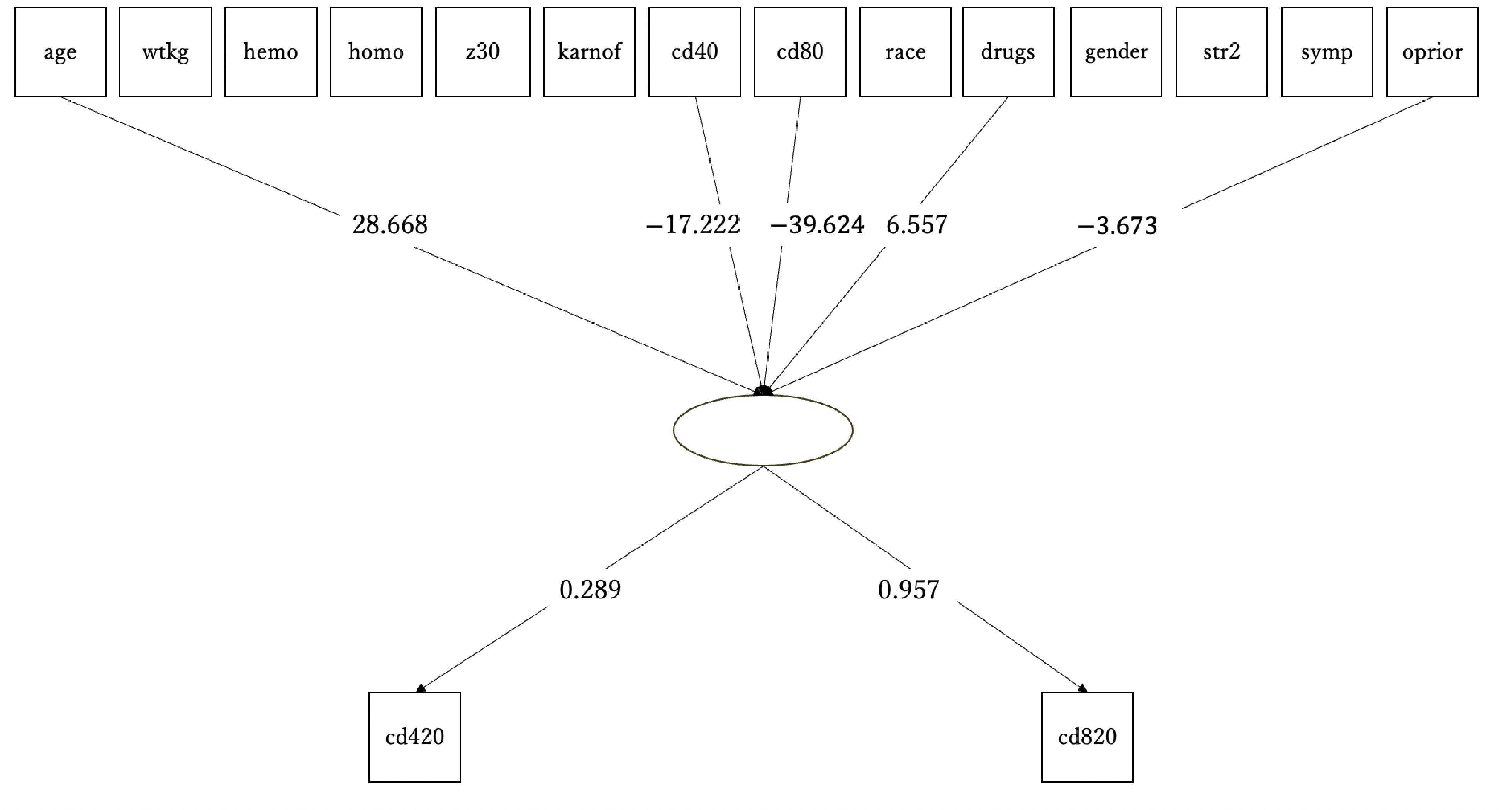}
\caption{Path diagram of the proposed method}
\label{ACTG_pass}
\end{figure}

\quad Here, we interpret the results obtained by applying the proposed method to the ACTG175 data.
We estimated the regression coefficient matrix for treatment effects with rank one.
Regularization parameters $\lambda$ and $\phi$ are determined through five-fold cross-validation. 
We visualize estimated regression coefficient matrices $\bm{W}$ and $\bm{V}$ in Figure \ref{ACTG_pass}. 
Based on the path diagram in Figure \ref{ACTG_pass}, we interpret the coefficients from the latent factor on both outcomes as positive effects.
Here, we show the covariates contributing to treatment effects on CD4 T and CD8 T cell counts. 
From the results, age and drugs have positive effects on CD4 T and CD8 T cell counts.
However, cd40, cd80, and oprior are negative effects on CD4 T and CD8 T cell counts.
As a result, we identified a subgroup in which the combination therapy showed superior efficacy to the monotherapy. 
The subgroup in which combination therapy showed efficacy for CD4 T cell count if the subjects are old, the baseline level of CD8T cells is low, the baseline level of CD8T cells is low includes subjects with a history of intravenous drug use and with no non-zidovudine antiretroviral therapy.
As for CD4T cell counts, the subgroup in which combination therapy showed efficacy for CD8 T cell count if the subjects are young, the baseline level of CD8T cells is high, the baseline level of CD8T cells is high includes subjects with no history of intravenous drug use and with non-zidovudine antiretroviral therapy.
Indeed, Fischl et al. (1995) 
highlighted the possibility of varying treatment effects across subgroups, particularly that treatment efficacy may be influenced by baseline cell counts. 
\section{Conclusions}
We propose a method for estimating HTE for multiple outcomes, which has not been sufficiently discussed in previous studies.
Specifically, a weighting method based on the robust sparse reduced-rank regression is proposed.
This method solves the vulnerability to outliers of multiple outcomes in the multivariate regression framework.
In addition, because the objective function of the proposed method is a least squares problem, it allows treatment effects estimation through a simple algorithm.
Furthermore, when true treatment effects have a low-rank structure existing methods might not correctly estimate the treatment effects
but we prove that, if an appropriate rank can be chosen, then true treatment effects could be estimated.
Through simulations, we have confirmed that the proposed method outperforms comparison methods in terms of MSE and Spearman's rank correlation in cases where the true coefficient matrix is on a low-rank space.
This shows that our method can estimate the treatment effects on multiple outcomes better than those of comparison methods and also identify subgroups for which the treatment is effective in some cases.
The proposed method was applied to ACTG175 data, and the results visualized. 
The visualization showed that it is possible to introduce latent factors for outcomes and select covariates related to treatment effects, identifying subgroups for multiple outcomes even when there are many outcomes and covariates.

\quad Finally, there are four areas for future work.
The first limitation of this study lies in the appropriate selection method for ranks.
In our simulation, ranks were selected by CV.
Therefore it takes much time to select the rank and the selection requires further discussions from a theoretical perspective.
The second is about theoretically guaranteeing that the treatment effects can be estimated when penalty terms are provided in the objective function. Penalty terms are incorporated to introduce the M-estimation framework and impose sparse constraints on the regression coefficient matrix representing the treatment effects in the objective function of the proposed method.
We need to further discuss the theoretical properties when including the penalty terms.
The third issue lies in determining the method for selecting initial values.
For both the simulations and real data, the initial values of the parameters were predetermined, but in future studies, a random start should be introduced to avoid local solutions.
The fourth limitation concerns the extension of data when the outcome is not continuous. 
We assume that outcomes are continuous data. However, in clinical trials, outcomes may be binary, such as disease-cure, or ordinal variables, such as pain severity. Therefore, it is necessary to generalize the objective function to deal with different outcomes and discuss this.


\vspace*{-8pt}


\appendix
\section{Proof of theorems and lemmas}
\begin{description}
    \item[\textbf{Proof of Lemma 1:}]
This is the same as the result of Theorem 1 in Yuki et al. (2023)
\item [\textbf{Proof of Lemma 2:}]
\begin{align*}
&E_{\bm{Y}^{(1)},\bm{Y}^{(-1)}}\left[\|\bm{Y}-\bm{X}^{*}_{\rm{(r.v.)}}\bm{\varGamma}\|_{F}^{2}|\bm{T}=\bm{I},\bm{X}_{\rm{(r.v.)}}=\bm{X}_{\rm{(o.b.)}}\right]\\
&+E_{\bm{Y}^{(1)},\bm{Y}^{(-1)}}\left[\|\bm{Y}+\bm{X}^{*}_{\rm{(r.v.)}}\bm{\varGamma}\|_{F}^{2}|\bm{T}=-\bm{I},\bm{X}_{\rm{(r.v.)}}=\bm{X}_{\rm{(o.b.)}}\right]\\
=&E_{\bm{Y}^{(1)},\bm{Y}^{(-1)}}[\|\bm{Y}^{(1)}-\bm{X}^{*}_{\rm{(r.v.)}}\bm{\varGamma}\|_{F}^{2}|\bm{X}_{\rm{(r.v.)}}=\bm{X}_{\rm{(o.b.)}}]\\
&+E_{\bm{Y}^{(1)},\bm{Y}^{(-1)}}\left[\|\bm{Y}^{(-1)}+\bm{X}^{*}_{\rm{(r.v.)}}\bm{\varGamma}\|_{F}^{2}|\bm{X}_{\rm{(r.v.)}}=\bm{X}_{\rm{(o.b.)}}\right]\\
=&E_{\bm{Y}^{(1)},\bm{Y}^{(-1)}}\left[{\rm{tr}}(\bm{Y}^{(1)\top}\bm{Y}^{(1)}-2\bm{Y}^{(1)}\bm{X}^{*}_{\rm{(r.v.)}}\bm{\varGamma}+\bm{\varGamma}^{\top}\bm{X}^{*\top}_{\rm{(r.v.)}}\bm{X}^{*}_{\rm{(r.v.)}}\bm{\varGamma})\right.
\\&\left.+{\rm{tr}}(\bm{Y}^{(-1)\top}\bm{Y}^{(-1)}+2\bm{Y}^{(-1)}\bm{X}^{*}_{\rm{(r.v.)}}\bm{\varGamma}+\bm{\varGamma}^{\top}\bm{X}^{*\top}_{\rm{(r.v.)}}\bm{X}^{*}_{\rm{(r.v.)}}\bm{\varGamma})|\bm{X}_{\rm{(r.v.)}}=\bm{X}_{\rm{(o.b.)}}\right]\\
=&E_{\bm{Y}^{(1)},\bm{Y}^{(-1)}}\left[{\rm{tr}}\left\{(\bm{Y}^{(1)\top}\bm{Y}^{(1)}+\bm{Y}^{(-1)\top}\bm{Y}^{(-1)})-2(\bm{Y}^{(1)}-\bm{Y}^{(-1)})\bm{X}^{*}_{\rm{(r.v.)}}\bm{\varGamma}\right.\right.\\
&\left.\left.+2\bm{\varGamma}^{\top}\bm{X}^{*\top}_{\rm{(r.v.)}}\bm{X}^{\top}_{\rm{(r.v.)}}\bm{\varGamma})\right\}|\bm{X}_{\rm{(r.v.)}}=\bm{X}_{\rm{(o.b.)}}\right]
\\=&E_{\bm{Y}^{(1)},\bm{Y}^{(-1)}}\left.\left[\left\|\frac{1}{\sqrt{2}}(\bm{Y}^{(1)}-\bm{Y}^{(-1)})-\sqrt{2}\bm{X}^{*}_{\rm{(r.v.)}}\bm{\varGamma}\right\|_{F}^{2}\right|\bm{X}_{\rm{(r.v.)}}=\bm{X}_{\rm{(o.b.)}}\right]\\
&+E_{\bm{Y}^{(1)},\bm{Y}^{(-1)}}\left[\left.{\rm{tr}}\left(\frac{1}{2}\bm{Y}^{(1)\top}\bm{Y}^{(1)}-\frac{3}{2}\bm{Y}^{(1)\top}\bm{Y}^{(-1)}+\frac{1}{2}\bm{Y}^{(-1)\top}\bm{Y}^{(-1)}\right)\right|\bm{X}_{\rm{(r.v.)}}=\bm{X}_{\rm{(o.b.)}}\right].
\end{align*}
Thus, we have proved the Lemma 2.
\item [\textbf{Proof of Theorem 1:}]
$\bm{L}_{w}(\bm{\varGamma}|\bm{X}_{{\rm{(r.v.)}}}=\bm{X}_{{\rm{(o.b.)}}})$ can be equationally transformed as follows;
\begin{align}
\label{wp_meidai_1}
&\bm{L}_{w}(\bm{\varGamma}|\bm{X}_{{\rm{(r.v.)}}}=\bm{X}_{{\rm{(o.b.)}}})
 \\ 
&=\left.E_{\bm{Y}}\left[\left\|\bm{Y}-\bm{X}_{\rm{(r.v.)}}^{*}\bm{\varGamma}\right\|_{F}^{2}\right|\bm{T}=\bm{I},\bm{X}_{\rm{(r.v.)}}=\bm{X}_{\rm{(o.b.)}}\right] \notag\\
\label{wp_meidai_2}
&\left.+E_{\bm{Y}}\left[\left\|\bm{Y}+\bm{X}_{\rm{(r.v.)}}^{*}\bm{\varGamma}\right\|_{F}^{2}\right|\bm{T}=-\bm{I},\bm{X}_{\rm{(r.v.)}}=\bm{X}_{\rm{(o.b.)}}\right]\\  
\label{wp_meidai_3}
&=E_{\bm{Y}^{(1)},\bm{Y}^{(-1)}}\left.\left[\left\|\frac{1}{\sqrt{2}}(\bm{Y}^{(1)}-\bm{Y}^{(-1)})-\sqrt{2}\bm{X}^{*}_{\rm{(r.v.)}}\bm{\varGamma}\right\|_{F}^{2}\right|\bm{X}_{\rm{(r.v.)}}=\bm{X}_{\rm{(o.b.)}}\right]\notag\\
&+E_{\bm{Y}^{(1)},\bm{Y}^{(-1)}}\left.\left[{\rm{tr}}\left(\frac{1}{2}\bm{Y}^{(1)\top}\bm{Y}^{(1)}-\frac{3}{2}\bm{Y}^{(1)\top}\bm{Y}^{(-1)}+\frac{1}{2}\bm{Y}^{(-1)\top}\bm{Y}^{(-1)}\right)\right|\bm{X}_{\rm{(r.v.)}}=\bm{X}_{\rm{(o.b.)}}\right],
\end{align}
where, we denote $=\bm{X}_{\rm{(r.v.)}}/2$ as $\bm{X}_{\rm{(r.v.)}}^{*}$ and identity matrix as
$\bm{I}\in\mathbb{R}^{n\times n}$.
Please see the Lemma 1 for the transformation of equation (\ref{wp_meidai_1}) to equation (\ref{wp_meidai_2}) and the Lemma 2 for the transformation of equation (\ref{wp_meidai_2}) to equation (\ref{wp_meidai_3}).
Therefore, $\bm{L}_{w}(\bm{\varGamma}|\bm{X}_{{\rm{(r.v.)}}}=\bm{X}_{{\rm{(o.b.)}}})$ is equivalent to solving the following minimisation problem;
\begin{align}
\label{wp_meidai_4}
&E_{\bm{Y}^{(1)},\bm{Y}^{(-1)}}\left.\left[\left\|\frac{1}{\sqrt{2}}(\bm{Y}^{(1)}-\bm{Y}^{(-1)})-\sqrt{2}\bm{X}^{*}_{\rm{(r.v.)}}\bm{\varGamma}\right\|_{F}^{2}\right|\bm{X}_{\rm{(r.v.)}}=\bm{X}_{\rm{(o.b.)}}\right].
\end{align}

\quad There exists $\hat{\bm{\varGamma}}$, which is the minimum solution of $\bm{L}_{w}(\bm{\varGamma}|\bm{X}_{{\rm{(r.v.)}}}=\bm{X}_{{\rm{(o.b.)}}})$ and we can denote 
\begin{equation}
\hat{\bm{\varGamma}}=\hat{\bm{W}}\hat{\bm{V}}^{\top},
\end{equation}
 using the matrix $\hat{\bm{W}}\in\mathbb{R}^{p\times r}$ and the column orthogonal matrix $\hat{\bm{V}}\in\mathbb{R}^{q\times r}$.
then, the $\hat{\bm{W}}$ that minimises the equation
\rm{(\ref{wp_meidai_4})} is denoted using $\hat{\bm{V}}$ as follows;
\begin{align*}
E_{\bm{Y}^{(1)},\bm{Y}^{(-1)}}\left[\hat{\bm{W}}\right]&=E_{\bm{Y}^{(1)},\bm{Y}^{(-1)}}\left[\frac{1}{2}(\bm{X}_{\rm{(o.b.)}}^{*\top}\bm{X}_{\rm{(o.b.)}}^{*})^{-1}\sqrt{2}\bm{X}_{\rm{(o.b.)}}^{*\top}\frac{1}{\sqrt{2}}(\bm{Y}^{(1)}-\bm{Y}^{(-1)})\hat{\bm{V}}\right]
\\&=E_{\bm{Y}^{(1)},\bm{Y}^{(-1)}}\left[2(\bm{X}_{\rm{(o.b.)}}^{\top}\bm{X}_{\rm{(o.b.)}})^{-1}\frac{1}{2}\bm{X}_{\rm{(o.b.)}}^{\top}(\bm{Y}^{(1)}-\bm{Y}^{(-1)})\hat{\bm{V}}\right]
\\&=E_{\bm{Y}^{(1)},\bm{Y}^{(-1)}}\left[(\bm{X}_{\rm{(o.b.)}}^{\top}\bm{X}_{\rm{(o.b.)}})^{-1}\bm{X}_{\rm{(o.b.)}}^{\top}\bm{X}_{\rm{(o.b.)}}\bm{\varGamma}_{\rm{(true)}}\hat{\bm{V}}\right]
\\&=E_{\bm{Y}^{(1)},\bm{Y}^{(-1)}}\left[\bm{\varGamma}_{\rm{(true)}}\hat{\bm{V}}\right],
\end{align*}
where $\bm{\varGamma}_{\rm{(true)}}\in\mathbb{R}^{p\times q}$ denotes the true regression coefficient matrix representing treatment effects.
Please see Theorem 2.2 in Reinsel and Velu (1998).
Therefore, the following equation;
\begin{align}
\label{wp_meidai_7}
\bm{X}_{\rm{(o.b.)}}\bm{W}^{*}=
\bm{X}_{\rm{(o.b.)}}\bm{\varGamma}_{\rm{(true)}}\bm{V}^{*}
\end{align}
holds, where $\bm{W}^{*}$ denote $E_{\bm{Y}^{(1)},\bm{Y}^{(-1)}}\left[\hat{\bm{W}}\right]$
and $\bm{V}^{*}$ denote $E_{\bm{Y}^{(1)},\bm{Y}^{(-1)}}\left[\hat{\bm{V}}\right]$.

\quad On the other hand, it follows that
\begin{align}
\label{wp_meidai8}
&E_{\bm{Y}^{(1)},\bm{Y}^{(-1)}}\left[\left.\bm{Y}^{(1)}\bm{V}^{*}-\bm{Y}^{(-1)}\bm{V}^{*}\right|\bm{X}_{\rm{(r.v.)}}=\bm{X}_{\rm{(o.b.)}}\right]\notag\\
=&E_{\bm{Y}^{(1)},\bm{Y}^{(-1)}}\left[\left(B(\bm{X}_{\rm{(r.v.)}})+\frac{1}{2}\bm{X}_{\rm{(r.v.)}}\bm{\varGamma}_{\rm{(true)}}+\bm{E}\right)\bm{V}^{*}\right. \notag
\\&\left.\left.-\left(B(\bm{X}_{\rm{(r.v.)}})-\frac{1}{2}\bm{X}_{\rm{(r.v.)}}\bm{\varGamma}_{\rm{(true)}}+\bm{E}\right)\bm{V}^{*}\right|\bm{X}_{\rm{(r.v.)}}=\bm{X}_{\rm{(o.b.)}}\right] \notag\\
=&\bm{X}_{\rm{(o.b.)}}\bm{\varGamma}_{\rm{(true)}}\bm{V}^{*}.
\end{align}
Therefore, From equation (\ref{wp_meidai_7}) and equation (\ref{wp_meidai8}), we find that 
\begin{align*}
&E_{\bm{Y}^{(1)},\bm{Y}^{(-1)}}\left[\left.\bm{Y}^{(1)}\bm{V}^{*}\bm{V}^{*\top}-\bm{Y}^{(-1)}\bm{V}^{*}\bm{V}^{*\top}\right|\bm{X}_{\rm{(r.v.)}}=\bm{X}_{\rm{(o.b.)}}\right]\\
&=\bm{X}_{\rm{(o.b.)}}{\bm{W}}^{*}\bm{V}^{*\top}.
\end{align*}
From the above, the theorem is proved.
\item [\textbf{Proof of Corollary 1:}]
If we assume that $\bm{\varGamma}_{\rm{(true)}}=\bm{\varGamma}^{*}\bm{V}^{*\top}$, the following equality transformation holds;

\begin{align}
\label{colo_1}
&E_{\bm{Y}^{(1)},\bm{Y}^{(-1)}}\left[\left.\bm{Y}^{(1)}\bm{V}^{*}\bm{V}^{*\top}-\bm{Y}^{(-1)}\bm{V}^{*}\bm{V}^{*\top}\right|\bm{X}_{\rm{(r.v.)}}=\bm{X}_{\rm{(o.b.)}}\right] \notag\\
&=E_{\bm{Y}^{(1)},\bm{Y}^{(-1)}}\left[\left.\left(B(\bm{X}_{(r.v.)})+\frac{1}{2}\bm{X}_{\rm{(r.v.)}}\bm{\varGamma}_{\rm{(true)}}+\bm
{E}\right)\bm{V}^{*}\bm{V}^{*\top}\right.\right. \notag
\\&-\left.\left.\left(B(\bm{X}_{(r.v.)})-\frac{1}{2}\bm{X}_{\rm{(r.v.)}}\bm{\varGamma}_{\rm{(true)}}+\bm
{E}\right)\bm{V}^{*}\bm{V}^{*\top}\right|\bm{X}_{\rm{(r.v.)}}=\bm{X}_{\rm{(o.b.)}}\right] \notag\\
&=E_{\bm{Y}^{(1)},\bm{Y}^{(-1)}}\left[\left.\bm{X}_{\rm{(r.v.)}}\bm{\varGamma}^{*}\bm{V}^{*\top}\bm{V}^{*}\bm{V}^{*\top}\right|\bm{X}_{\rm{(r.v.)}}=\bm{X}_{\rm{(o.b.)}}\right] \notag\\
&=\bm{X}_{\rm{(o.b.)}}\bm{\varGamma}_{\rm{(true)}}  \notag\\
&=E\left[\bm{Y}^{(1)}-\bm{Y}^{(-1)}|\bm{X}_{\rm{(r.v.)}}=\bm{X}_{\rm{(o.b.)}}\right].
\end{align}
Here, from Theorem 1, the following equation holds:
\begin{align}
\label{colo_2}
&E_{\bm{Y}^{(1)},\bm{Y}^{(-1)}}\left[\left.\bm{Y}^{(1)}\bm{V}^{*}\bm{V}^{*\top}-\bm{Y}^{(-1)}\bm{V}^{*}\bm{V}^{*\top}\right|\bm{X}_{\rm{(r.v.)}}=\bm{X}_{\rm{(o.b.)}}\right] \notag\\
&=\bm{X}_{\rm{(o.b.)}}{\bm{W}}^{*}\bm{V}^{*\top},
\end{align}
where the matrix $\hat{\bm{W}}\in\mathbb{R}^{p\times r}$ and the column orthogonal matrix $\hat{\bm{V}}
\in\mathbb{R}^{q\times r}$ are satisfying  $\hat{\bm{\varGamma}}=\hat{\bm{W}}\hat{\bm{V}}^{\top}$, $\hat{\bm{W}}^{*}=E_{\bm{Y}^{(1)},\bm{Y}^{(-1)}}\left[\hat{\bm{W}}\right]$, and $\hat{\bm{V}}^{*}=E_{\bm{Y}^{(1)},\bm{Y}^{(-1)}}\left[\hat{\bm{V}}\right]$.
\end{description}

\section{Figures pertaining to simulation study}
In this session, we show the results of the numerical simulations (not shown in the paper) are presented below.

\begin{figure}[H]
    \includegraphics[keepaspectratio, scale=0.9]
    {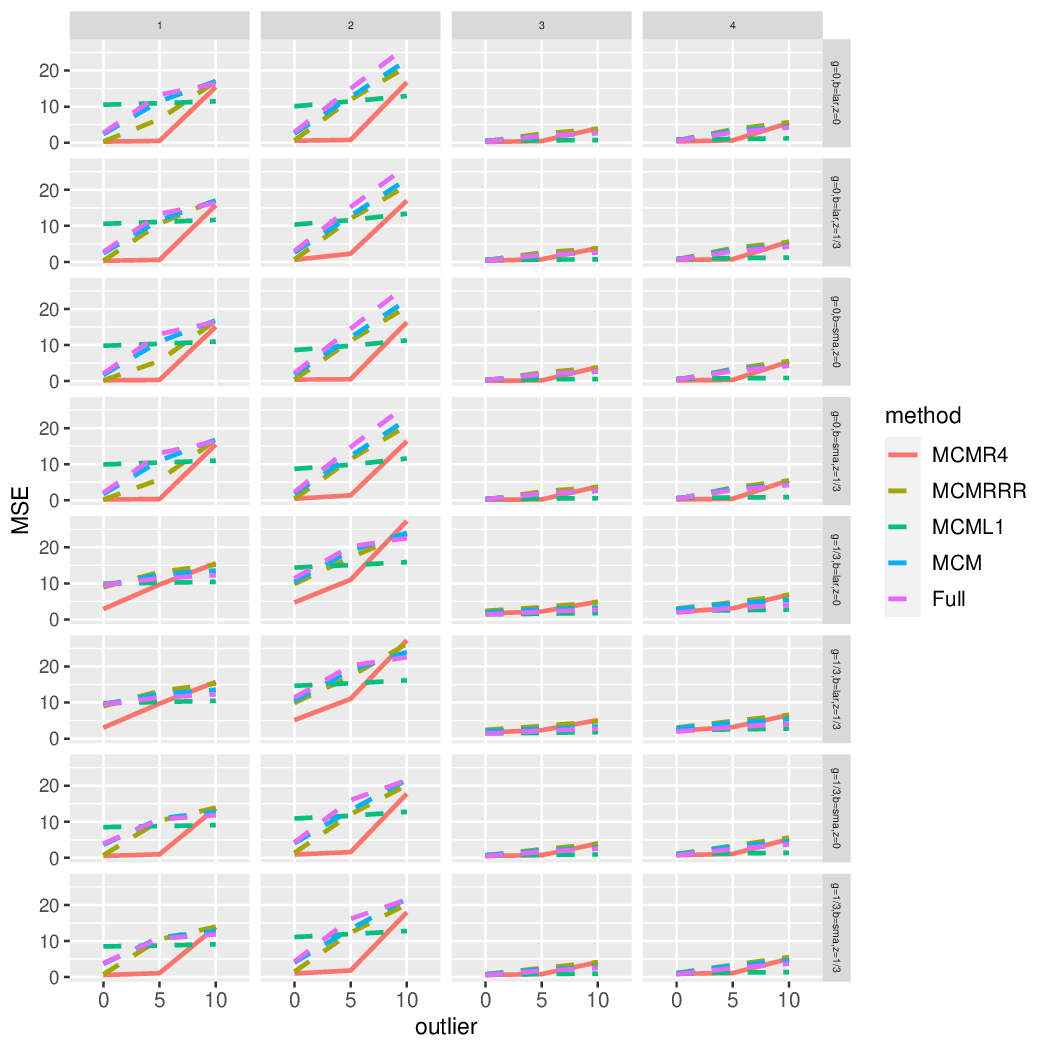}
      \centering
    \caption{\small{Transition plot of MSE for percentage of outliers ($p=$10/  RCT)}}
    \label{sim_mse5}
\end{figure}
\newpage

\begin{figure}[H]
    \includegraphics[keepaspectratio, scale=0.9]
    {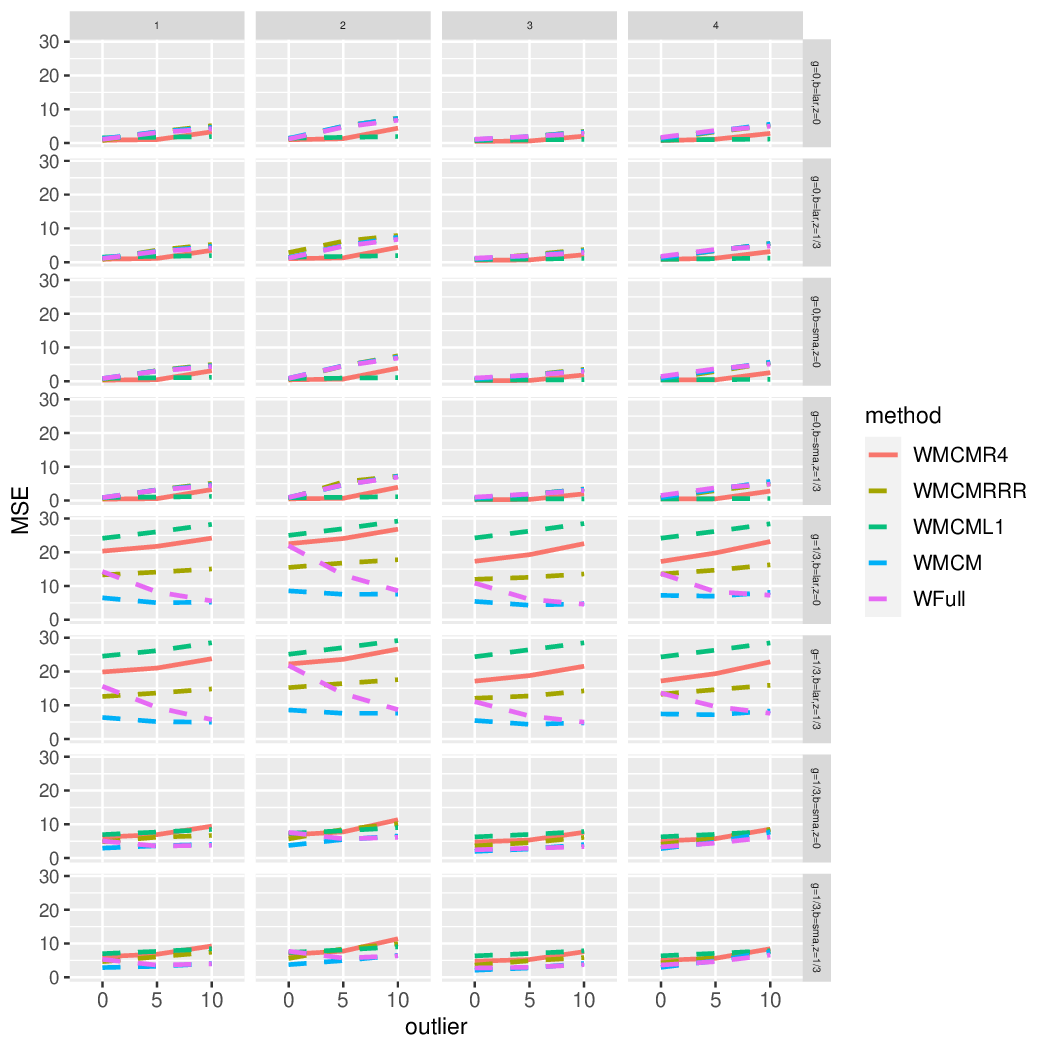}
      \centering
    \caption{\small{Transition plot of MSE for percentage of outliers ($p=$10/    Observed studies)}}
    \label{sim_mse13}
\end{figure}

\begin{figure}[H]
    \includegraphics[keepaspectratio, scale=0.9]
    {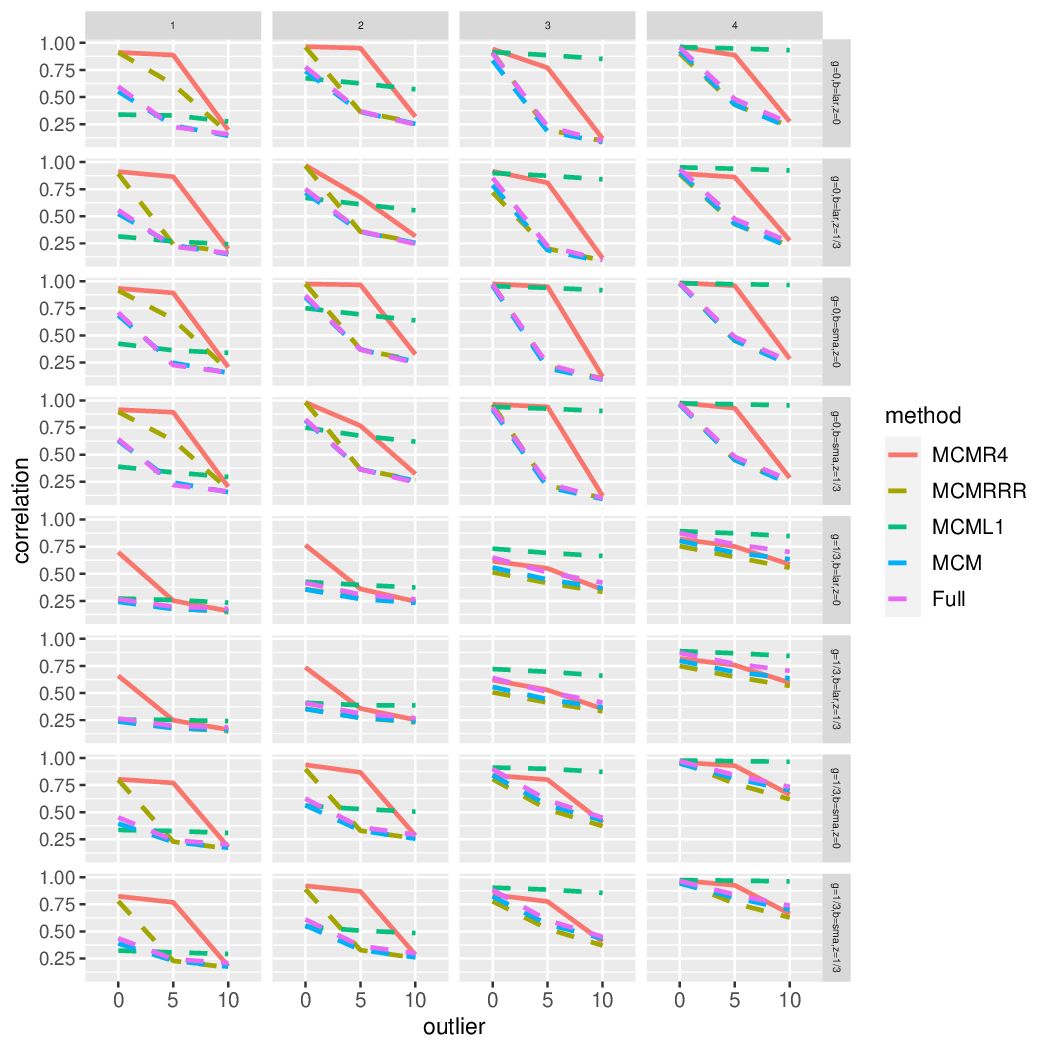}
      \centering
   \caption{\small{Transition plot of rank correlation for percentage of outliers ($p=$10/  RCT)}}
    \label{sim_spia5}
    
\end{figure}
\begin{figure}[H]
    \includegraphics[keepaspectratio, scale=0.9]
    {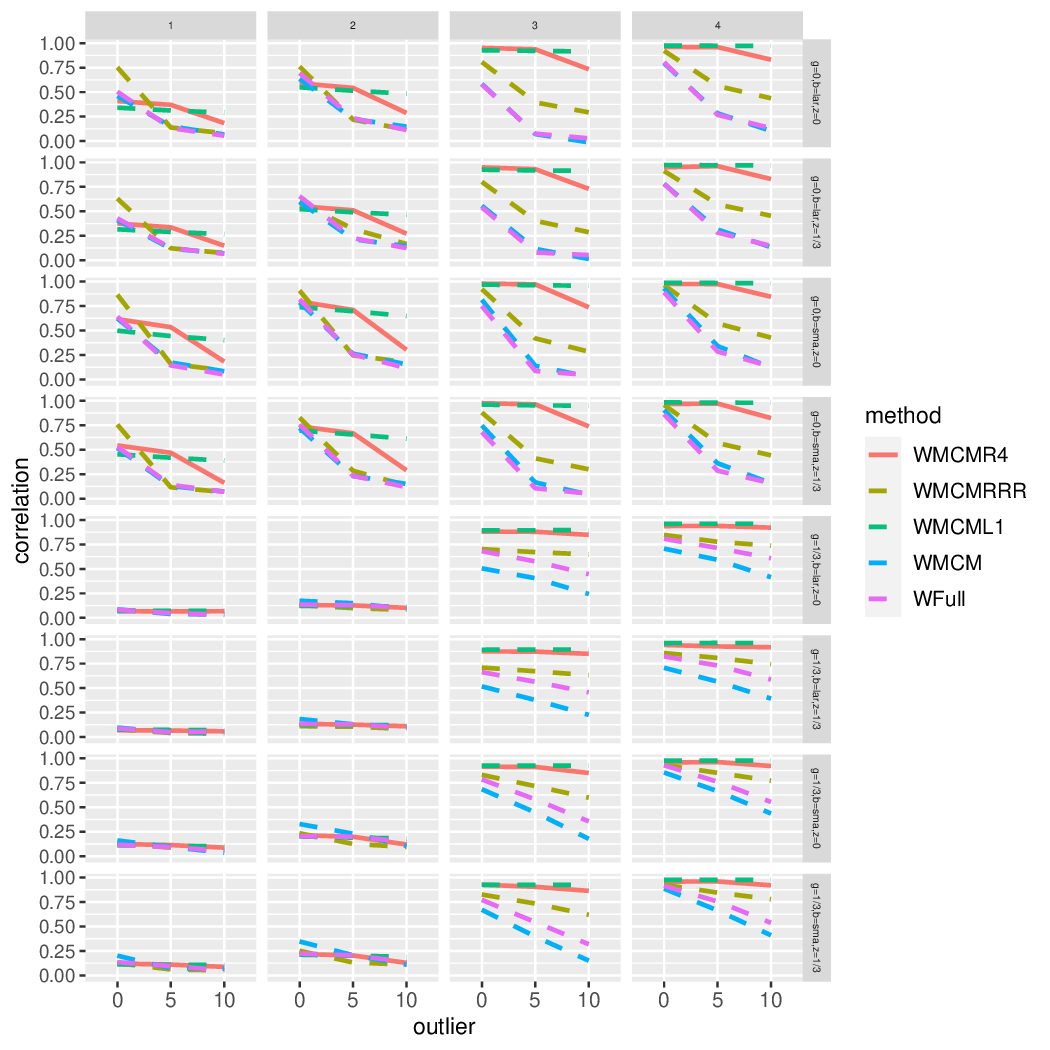}
      \centering
    \caption{\small{Transition plot of rank correlation for percentage of outliers ($p=$10/  Observed studies)}}
    \label{sim_spia6}
\end{figure}

\begin{figure}[H]
    \includegraphics[keepaspectratio, scale=0.9]
    {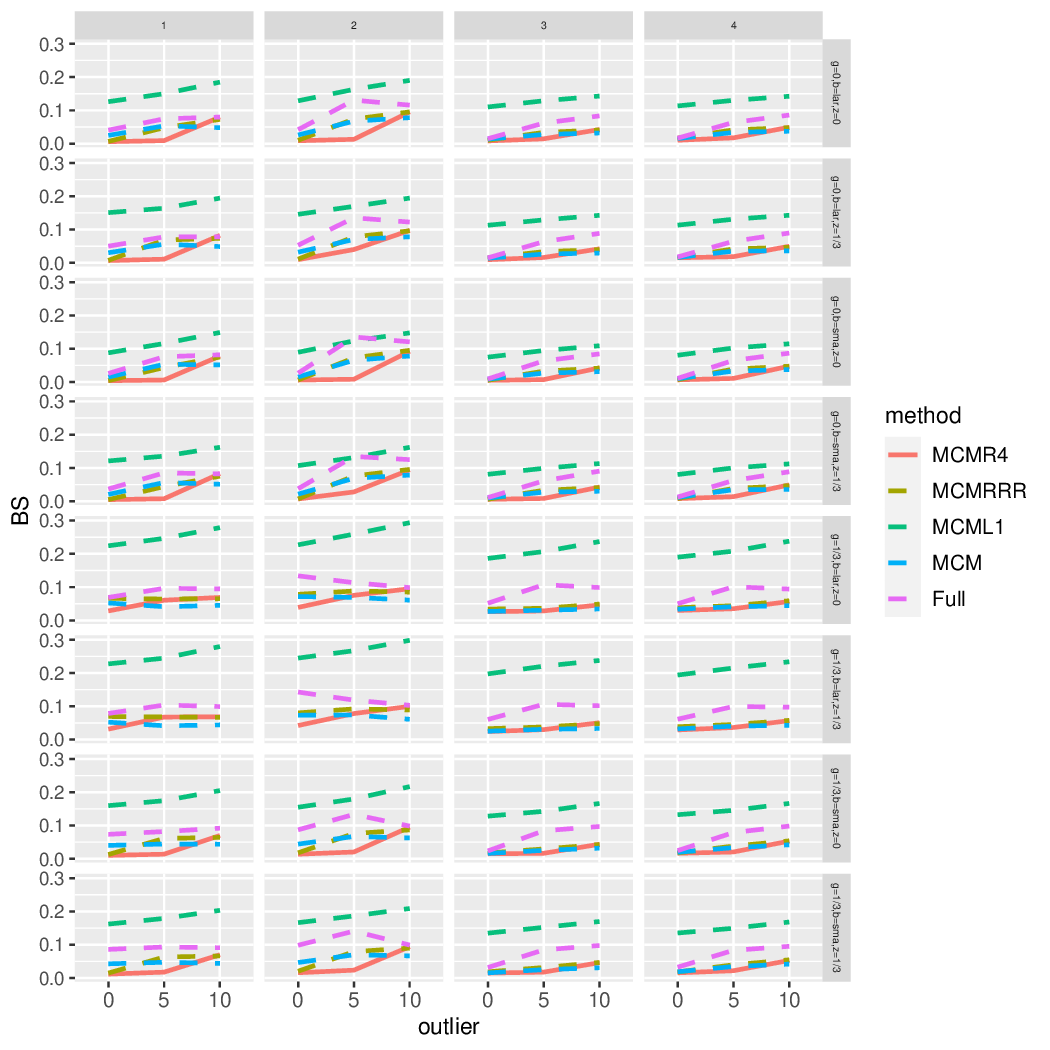}
      \centering
    \caption{\small{Transition plot of bias for percentage of outliers ($p=$50/  RCT)}}
    \label{sim_BS5}
    
\end{figure}
\begin{figure}[H]
    \includegraphics[keepaspectratio, scale=0.9]
    {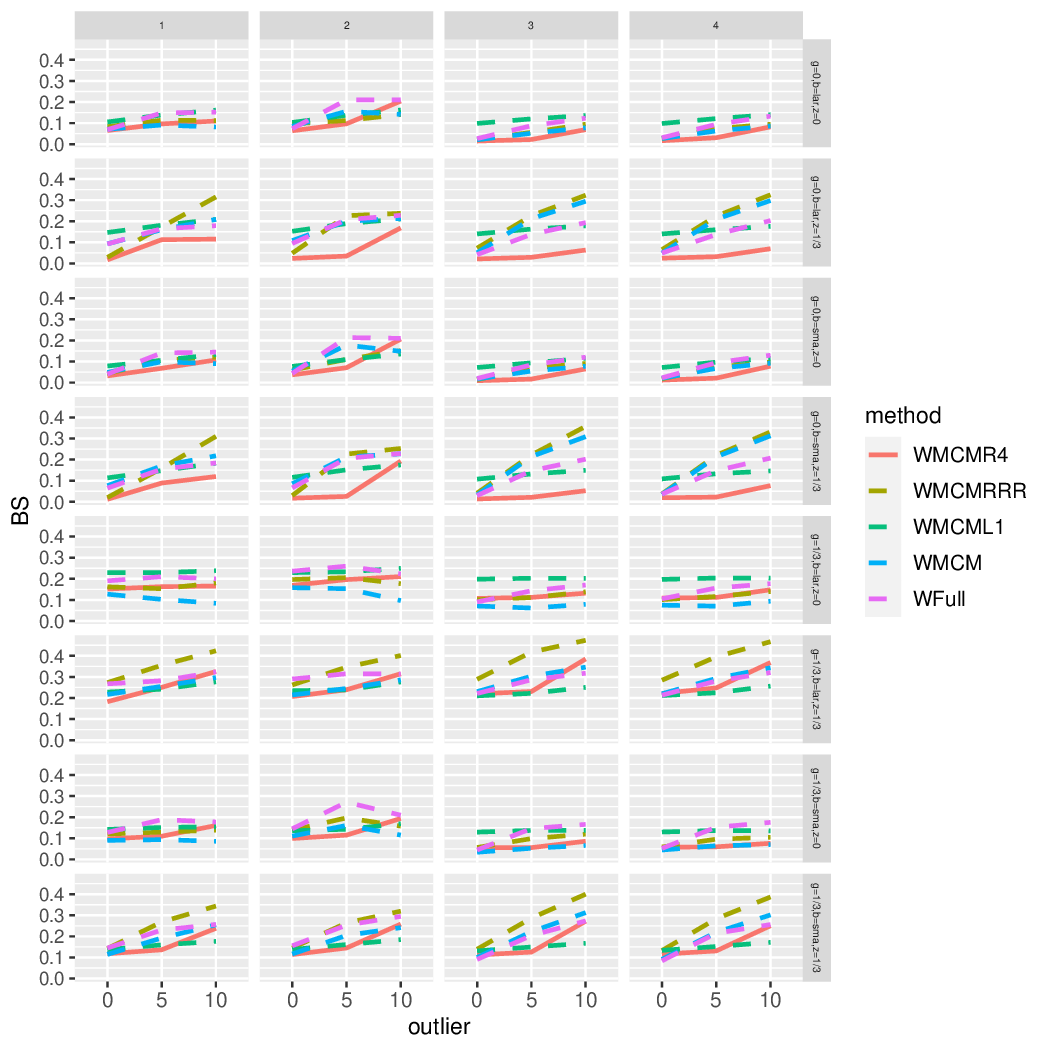}
      \centering
    \caption{\small{Transition plot of bias for percentage of outliers ($p=$50/  Observed studies)}}
    \label{sim_BS6}
\end{figure}    

\begin{figure}[H]
    \includegraphics[keepaspectratio, scale=0.9]
    {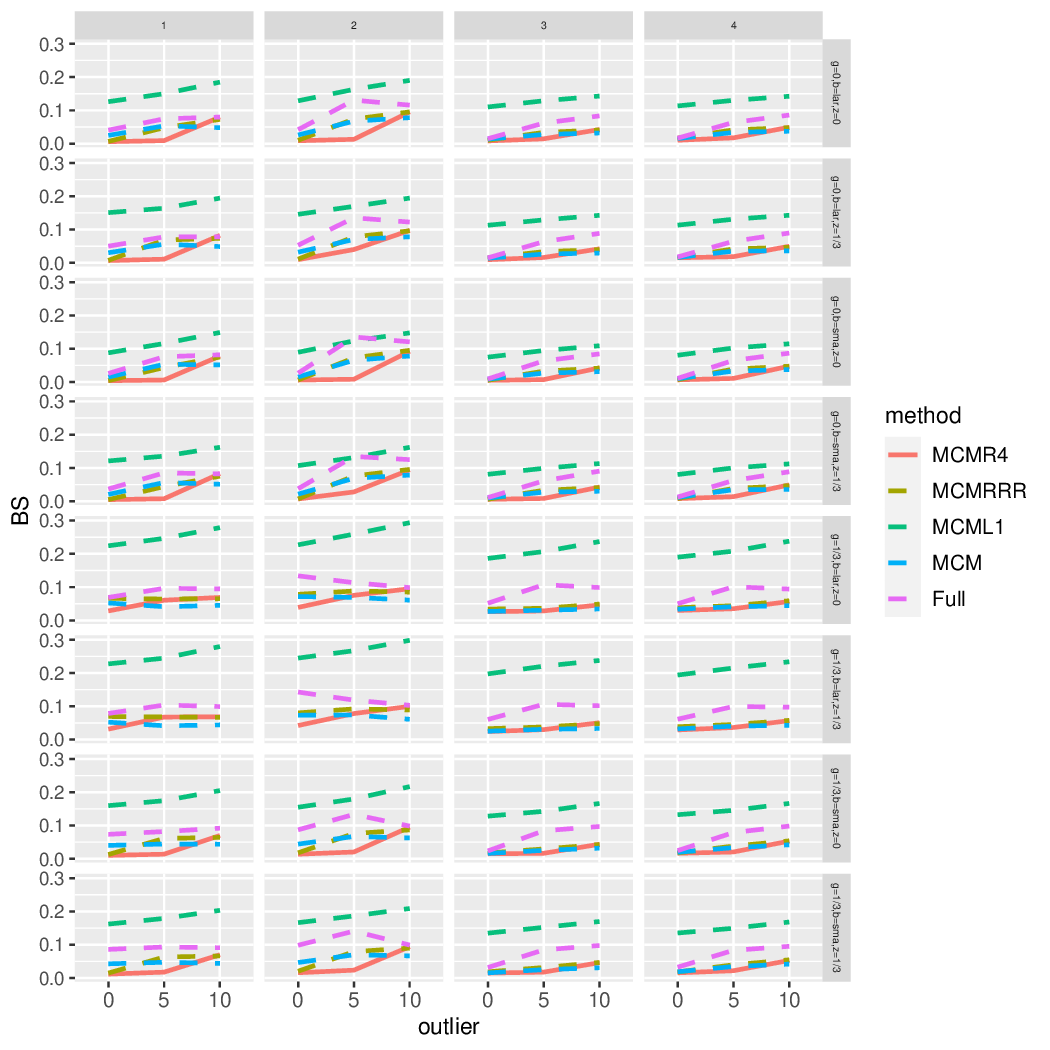}
      \centering
    \caption{\small{Transition plot of bias for percentage of outliers ($p=$10/  RCT)}}
    \label{sim_BS7}
    
\end{figure}
\begin{figure}[H]
    \includegraphics[keepaspectratio, scale=0.9]
    {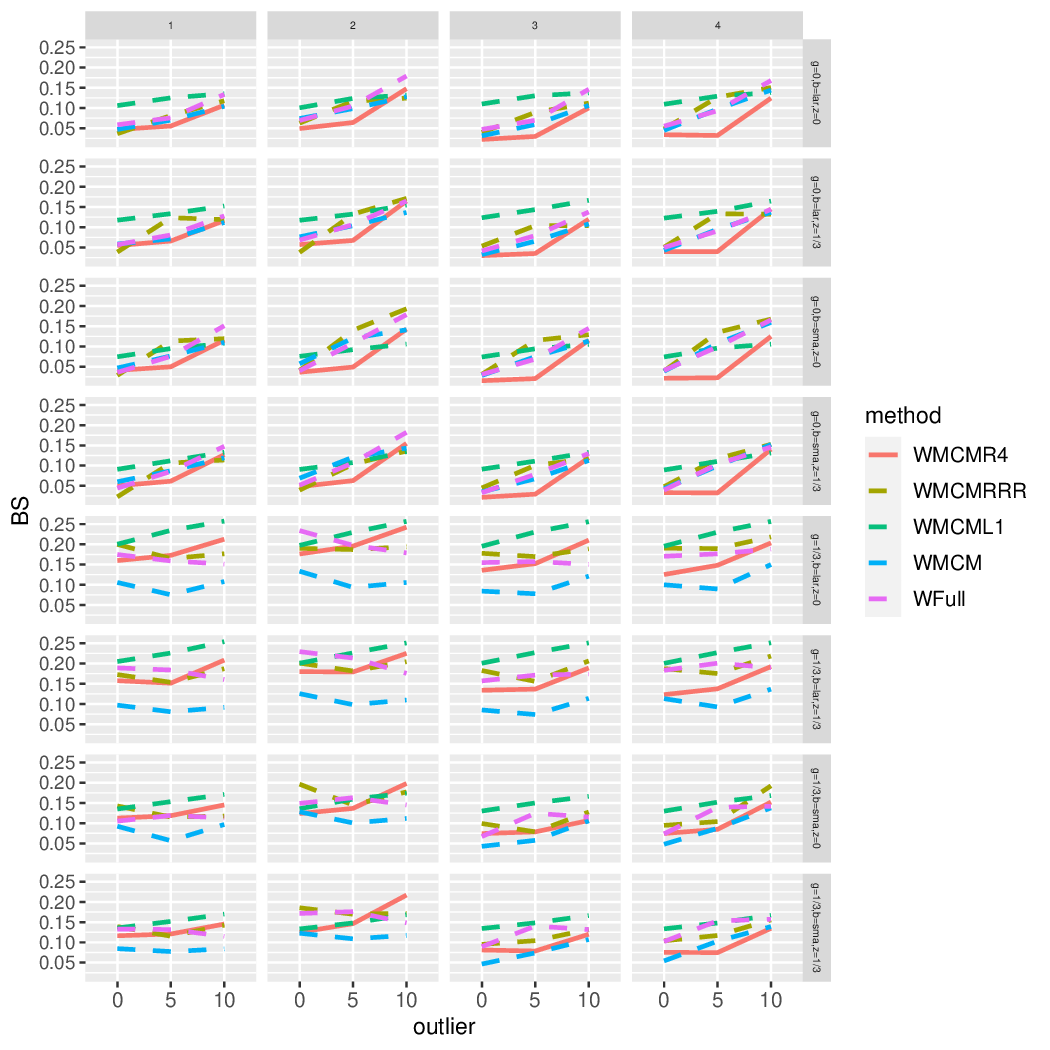}
      \centering
    \caption{\small{Transition plot of bias for percentage of outliers ($p=$10/  Observed studies)}}
    \label{sim_BS8}
\end{figure}    

\begin{figure}[H]
    \includegraphics[keepaspectratio, scale=0.9]
    {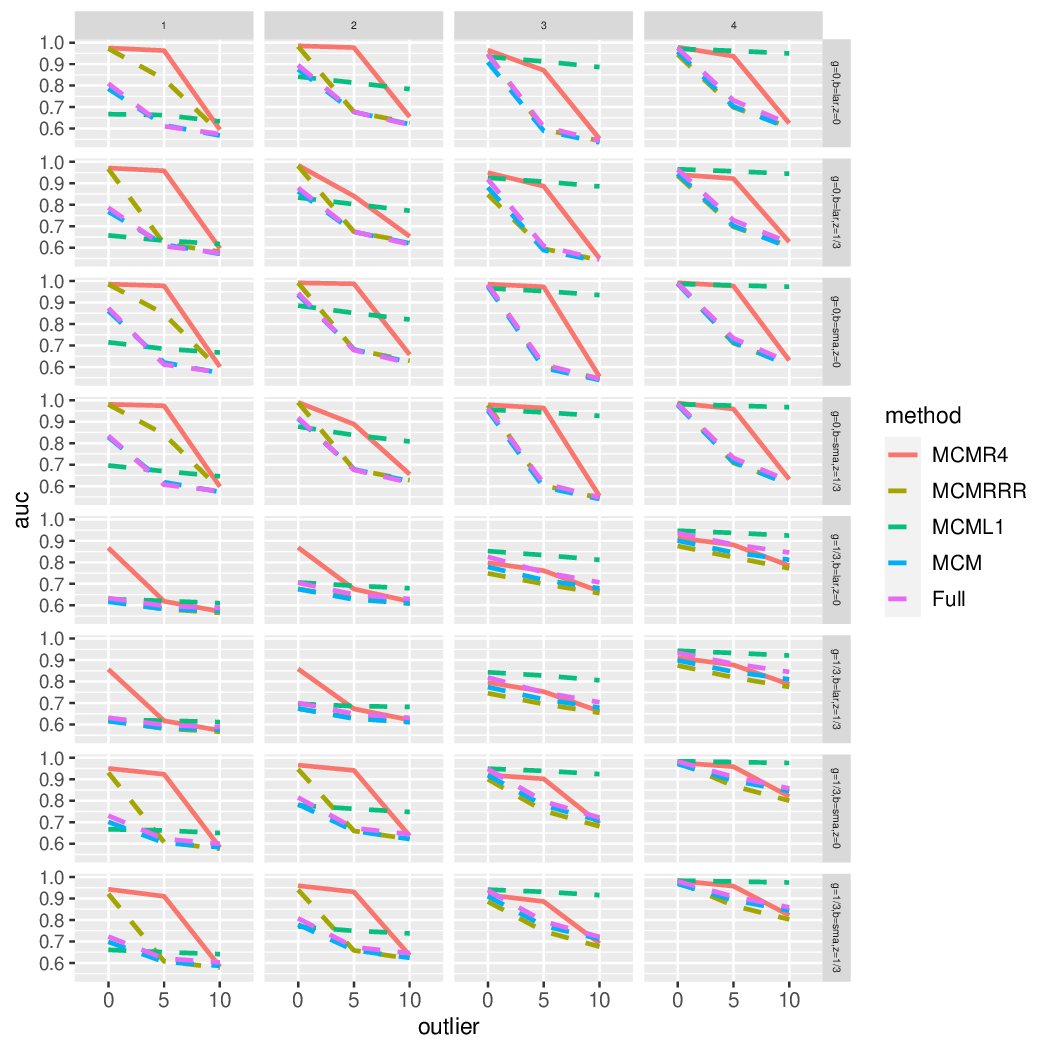}
      \centering
   \caption{\small{Transition plot of AUC for percentage of outliers ($p=$50/  RCT)}}
    \label{sim_AUC5}
    
\end{figure}
\begin{figure}[H]
    \includegraphics[keepaspectratio, scale=0.9]
    {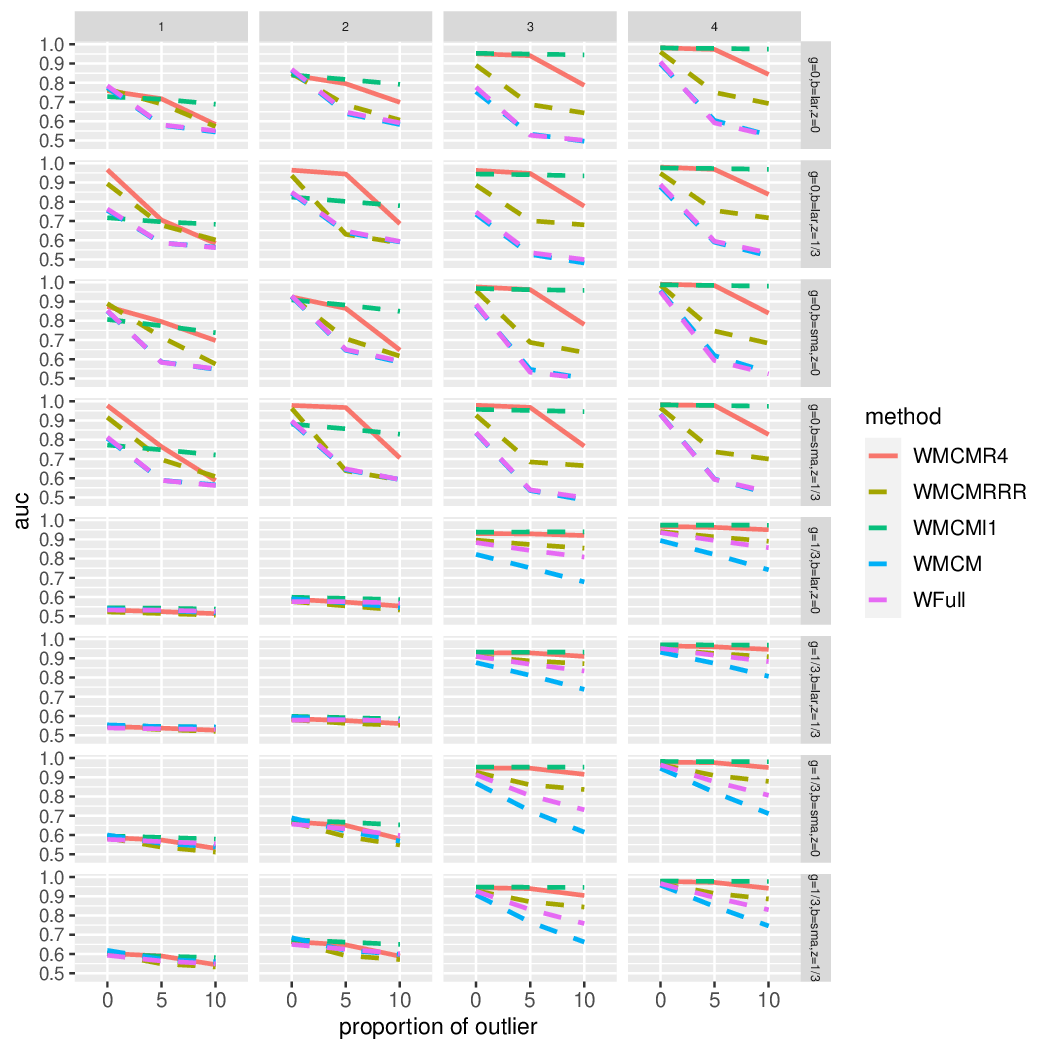}
      \centering
    \caption{\small{Transition plot of AUC for percentage of outliers ($p=$50/  Observed studies)}}
    \label{sim_AUC6}
\end{figure}    

\begin{figure}[H]
    \includegraphics[keepaspectratio, scale=0.9]
    {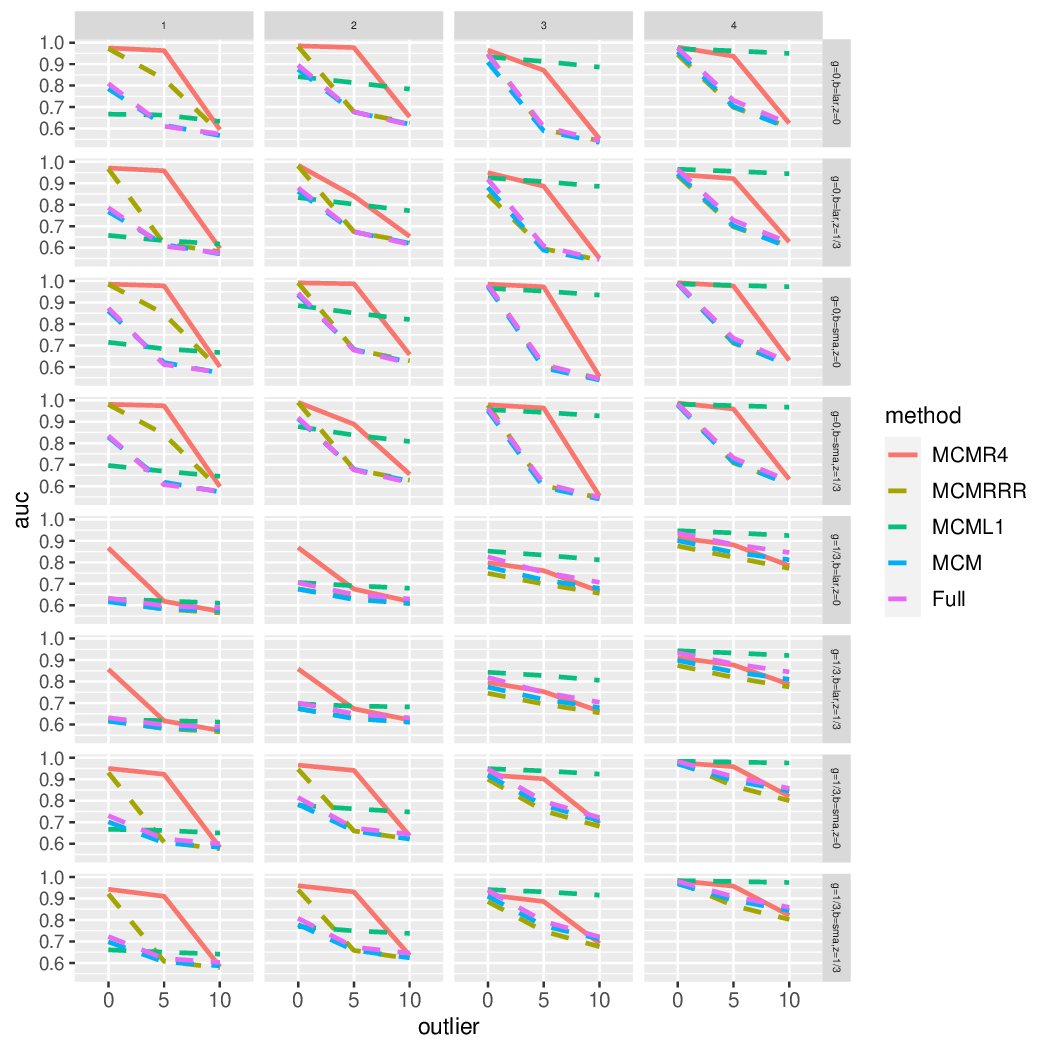}
      \centering
    \caption{\small{Transition plot of AUC for percentage of outliers ($p=$10/  RCT)}}
    \label{sim_AUC7}
    
\end{figure}
\begin{figure}[H]
    \includegraphics[keepaspectratio, scale=0.9]
    {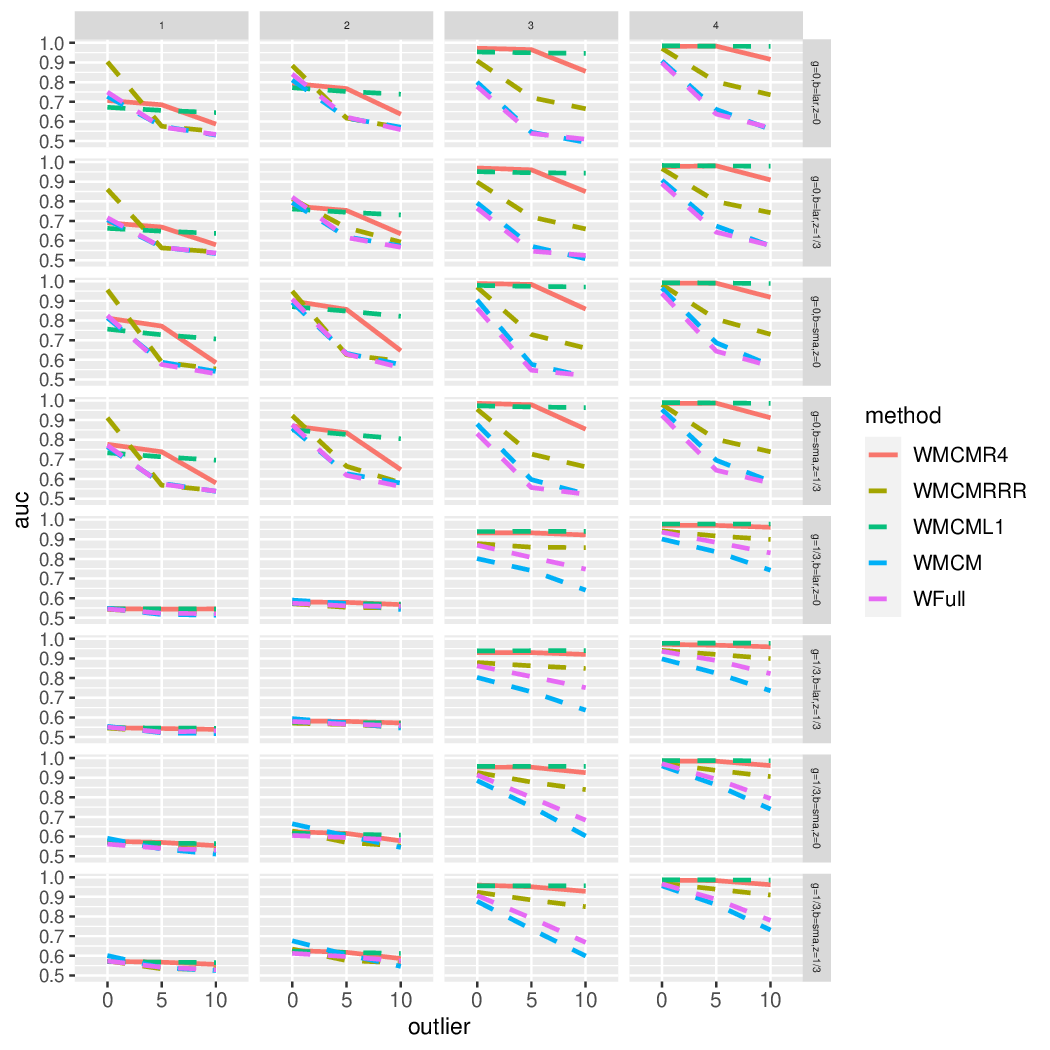}
      \centering
    \caption{\small{Transition plot of AUC for percentage of outliers ($p=$10/  Observed studies)}}
    \label{sim_AUC(}
\end{figure}

\end{document}